\documentclass[11pt,a4paper]{article}
\usepackage{jcappub}

\pdfoutput=1

\usepackage{tabularx}
\usepackage{amsmath,amssymb}
\usepackage{appendix}
\usepackage{graphicx}
\usepackage{color}
\usepackage{xcolor}
\usepackage{cancel}
\usepackage{subfigure}
\usepackage{CJKutf8}
\usepackage{multirow}
\usepackage{amssymb,changes}

\usepackage{booktabs} 

\usepackage{hyperref}

\usepackage{color}

\usepackage{bbding}

\newcommand{\Rmnum}[1]{\expandafter\@slowromancap\romannumeral #1@}
\makeatother

\newcommand{\tc}{\textcolor{black}}

\title{Pseudo-Goldstone Dark Matter from Primordial Black Holes: Gravitational Wave Signatures and Implications for KM3-230213A Event at KM3NeT}
\author[a]{Siyu Jiang}
\affiliation[a]{MOE Key Laboratory of TianQin Mission, TianQin Research Center for
Gravitational Physics \& School of Physics and Astronomy, Frontiers
Science Center for TianQin, Gravitational Wave Research Center of CNSA, 
Sun Yat-sen University (Zhuhai Campus), Zhuhai 519082, China}
\emailAdd{jiangsy36@mail2.sysu.edu.cn}

\author[a]{Fa Peng Huang\footnote{Corresponding author.}}
\emailAdd{huangfp8@sysu.edu.cn}

\abstract{
In many well-motivated new physics models, the pseudo-Nambu-Goldstone boson (pNGB) from $U(1)$ symmetry breaking emerges as a promising dark matter candidate. Its coupling, suppressed by the symmetry breaking scale, prevents thermal equilibrium in the early Universe for high scale symmetry breaking. Thus, pNGB dark matter is predominantly produced via non-thermal mechanisms, such as the freeze-in process through a new portal coupling.
In this work, we explore a novel mechanism for the production of pNGB dark matter even with feeble Higgs portal coupling—arising from Hawking radiation or superradiance of primordial black holes. We systematically investigate the production of light and heavy pNGB dark matter, both for Schwarzschild and Kerr black holes. We also discuss its potential gravitational wave signatures from domain wall collapse, density perturbations, and Hawking radiation. If the ultraviolet (UV) model is considered, the recent $\mathcal{O}$~(100) PeV neutrino event KM3-230213A at KM3NeT can be naturally explained.}

\keywords{dark matter theory, primordial black holes, gravitational waves/sources}

\begin{document}
\begin{CJK}{UTF8}{gbsn}
	
\maketitle
		
\section{Introduction}
Exploring the production mechanism and the microscopic nature of dark matter (DM) remains a central challenge in both cosmology and particle physics. The Standard Model (SM) of particle physics lacks viable DM candidates, necessitating the introduction of new physics beyond the SM. The astrophysical observations, direct detection experiments, and collider searches have placed stringent constraints on fundamental DM properties, such as its mass and interaction cross section. The traditional DM candidate, Weakly Interacting Massive Particles (WIMPs), produced via thermal freeze-out, is increasingly disfavored by direct detection results~\cite{SuperCDMS:2015eex,LUX:2016ggv,PandaX-II:2016vec,CRESST:2015txj,PICO:2017tgi}. 
This situation motivates the exploration of alternative DM candidates and novel production mechanisms.

The pseudo Nambu-Goldstone boson (pNGB) DM associated with U(1) symmetry breaking can naturally evade the severe constraints from direct detection~\cite{Gross:2017dan,Azevedo:2018exj,Ishiwata:2018sdi,Huitu:2018gbc,Arina:2019tib}. For example, in the case of lepton number symmetry $U(1)_{B-L}$ breaking, the right-handed neutrino gains heavy Majorana mass which can explain the smallness of the neutrino mass via the seesaw mechanism.  The Majoron, as the 
massless Nambu-Goldstone boson (NGB),
results in this case. The mass of pNGB appears when introducing an explicit U(1) breaking term which could come from the effect of quantum gravity~\cite{Rothstein:1992rh}, coupling with another scalar~\cite{Abe:2020iph,Okada:2020zxo,Mohapatra:2023aei,Liu:2022evb,Chao:2022blc}, neutrino Dirac Yukawa coupling~\cite{Frigerio:2011in}, and so on.

Due to the nature of NGB, all couplings of pNGB DM are inversely proportional to the vacuum expectation value (VEV) associated with the U(1) symmetry breaking. Therefore, from this viewpoint, if the breaking scale is high, it seems preferred for pNGB DM to be produced by freeze-in mechanism~\cite{Hall:2009bx} instead of freeze-out. The DM is assumed to be never in thermal equilibrium with the SM particles. Especially for some TeV-scale and heavier pNGB DM, for which the VEV has to be large enough in order for the DM to be long-lived, it is important to investigate its production mechanism. 

Primordial black holes (PBHs), as hypothetical macroscopic objects, may form due to density perturbations following the end of inflation. If a PBH's mass is less than $10^9~\mathrm{g}$, it would completely evaporate through Hawking radiation before Big Bang nucleosynthesis (BBN). In principle, black holes can efficiently produce any particles with masses below their surface temperature. These particles, with masses significantly exceeding the temperature of the Universe, naturally decouple from thermal equilibrium. This provides a novel mechanism for DM production~\cite{Allahverdi:2017sks,Lennon:2017tqq,Boucenna:2017ghj,Morrison:2018xla,Hooper:2019gtx,Adamek:2019gns,Gondolo:2020uqv,Chaudhuri:2020wjo,Masina:2020xhk,Baldes:2020nuv,Bernal:2020kse,Bernal:2020bjf,Bernal:2021bbv,Bernal:2021yyb,Barman:2021ost,Samanta:2021mdm,Chen:2023lnj,Kim:2023ixo,Shallue:2024hqe,Carr:2020mqm}, and matter-antimatter asymmetry~\cite{Baumann:2007yr,Hook:2014mla,Fujita:2014hha,Hooper:2020otu,Perez-Gonzalez:2020vnz,Datta:2020bht,Bernal:2022pue,Ambrosone:2021lsx,Hamada:2016jnq}. 
PBHs may also exhibit memory effects, altering their evolution and impacting DM production~\cite{Haque:2024eyh,Barman:2024iht,Borah:2024bcr,Basumatary:2024uwo,Zantedeschi:2024ram}, matter-antimatter asymmetry~\cite{Barman:2024kfj}, and gravitational wave (GW) signals~\cite{Bhaumik:2024qzd,Barman:2024ufm,Loc:2024qbz}. In recent years, the production of superheavy DM through superradiance has also attracted considerable attention~\cite{Bernal:2022oha,Ghoshal:2023fno,Manno:2025dhw,Athron:2024fcj}.

In this work, we systematically investigate the production of pNGB DM from Schwarzs-child and Kerr black holes, considering the effects of Hawking evaporation and superradiance. Furthermore, 
this DM production mechanism, along with the pNGB DM model, gives rise to a variety of GW signals: 
1. GWs from the annihilation of domain walls (DWs) arising from $Z_2$ symmetry breaking.  
2. Induced GWs generated by density perturbations, as PBHs may transiently dominate the Universe before their decay.  
3. High-frequency GWs directly emitted from PBHs. 
Moreover, if pNGB DM mass is around $\sim 440~\mathrm{PeV}$, its decay, facilitated by an ultraviolet (UV) extension of the model, could produce high-energy neutrinos, potentially explaining the KM3-230213A event observed at KM3NeT.

The paper is organized as follows. In section~\ref{model}, we briefly review the simple pNGB DM model. We review the Hawking radiation of Schwarzschild black holes, the freeze-in production, and discuss the light and heavy pNGB DM separately in section~\ref{SchDM}. In section \ref{KerDM}, we investigate the production of pNGB DM from Kerr black holes, including Hawking radiation and superradiance. The GW signals from DW collapse, density perturbations and the PBH evaporation are given in section~\ref{GWs}. Possible interpretation of the neutrino event of KM3-230213A is shown in section~\ref{KM3}. Finally, the conclusions and discussions are given in section~\ref{conclusions}.

\section{pNGB model}\label{model}
Firstly, we briefly review the minimal pNGB model. The scalar field $S$ possesses a dark $U(1)_D$ symmetry. The tree-level potential about SM Higgs $H$ and $S$ is
\begin{equation}\label{potential}
	V(H, S)=-\frac{\mu_{H}^2}{2}|H|^2+\frac{\lambda_{H}}{2}|H|^4 -\frac{\mu_{S}^2}{2}|S|^2+\frac{\lambda_{S}}{2}|S|^4+ \lambda_{HS} |H|^2 |S|^2-\frac{m^2}{4}\left(S^2+S^{* 2}\right)\,\,,
\end{equation}
where the first two terms are the potential of the SM Higgs. The term proportional to $m^2$ is the explicit breaking term, which gives mass to the pNGB 
$\chi$. The mass of $\chi$ can be superheavy, $M_\chi > 100~\mathrm{TeV}$, like in the UV model in Refs.~\cite{Abe:2020iph,Okada:2020zxo,Mohapatra:2023aei,Liu:2022evb} where the explicit breaking term can be derived by higher symmetry breaking. And in this case the pNGB DM could decay into neutrinos through scalar mixing or active-sterile neutrino mixing. The signals may interpret the results reported by KM3NeT recently and we will discuss later.
After introducing the explicit term, the $U(1)_D$ symmetry is reduced to $Z_2$ symmetry between $S \rightarrow -S$.
There also exists a CP symmetry as $S \rightarrow S^*$ in the potential.
After the symmetry breaking, the two scalar fields $H$ and $S$ can be written as
\begin{equation}
	H=\frac{1}{\sqrt{2}}\binom{0}{v+h}, \quad S=\frac{v_s+s}{\sqrt{2}} e^{i \chi / v_s}\,\,,
\end{equation}
where we have used the unitary gauge to eliminate the would-be NGBs in SM Higgs doublet $H$. 

If the mixing coupling $\lambda_{HS}$ is nonzero, $h$ and $s$ should be diagonalized. 
However, since we assume $v_s \gg v$, one can ignore the mixing, and just take $h$ and $s$ as physical eigenstate.
The stationary conditions of the VEVs are solved as $\mu_{S}^2=\lambda_{S} v_s^2-m^2$, and after the $U(1)_D$ breaking, the masses of $s$ and $\chi$ are given by
\begin{equation}
	M_s^2=\lambda_{S} v_s^2, \quad M_\chi^2=m^2\,\,.
\end{equation}
We require $\lambda_{HS}, \lambda_S \lesssim 10^{-6}$ so that the dark particles do not come into equilibrium. Then we get $M_s \leq 10^{-3}M_{\mathrm{Pl}}$ for $v_s \leq M_{\mathrm{Pl}}$ where $M_{\mathrm{Pl}} = 1/\sqrt{G_N} \simeq 1.22\times 10^{19}~\mathrm{GeV}$ is the Planck mass.
Because the vacuum value is real, the CP symmetry is unbroken then the Lagrangian is invariant under $\chi \rightarrow -\chi$, so pNGB is stable and can serve as the DM candidate.

It is worth noticing that when the $U(1)_D$ coincides with the $B-L$ symmetry, the complex singlet $S$ can give Majorana mass to the right-handed neutrinos by introducing the interaction term $S \overline{\nu_{R}^c} \nu_{R}$, through which the traditional seesaw mechanism works. In this case, the pNGB is called Majoron and it can decay into neutrinos through active-sterile neutrino mixing, the decay width reads~\cite{Manna:2022gwn},
\begin{equation}\label{decaywidth}
	\begin{aligned}
		\Gamma_{\chi \rightarrow \nu \nu}  \simeq \frac{1}{10^{19} ~\mathrm{s}}\left(\frac{M_\chi}{1~\mathrm{TeV}}\right)\left(\frac{8 \times 10^{11}~\mathrm{GeV}}{v_s}\right)^2\left(\frac{\sum_j m_{\nu_j}^2}{2.6 \times 10^{-3}~\mathrm{eV}^2}\right)\,\,.
	\end{aligned}
\end{equation}
The expression here uses the best fit values of atmospheric and solar mass-splittings~\cite{deSalas:2020pgw}. It can be seen that in order to make the Majoron be long-lived, the VEV $v_\phi$ has to be large enough. In order for Majoron to be a viable DM candidate, its lifetime should be larger than the lifetime of the Universe, $\Gamma_{\chi \rightarrow \nu \nu}^{-1}>\tau_U \sim \mathcal{O}\left(10^{19}\right) \sec$. For MeV Majoron, the symmetry breaking scale is limited to be larger than $\mathcal{O}(10^{11})~\mathrm{GeV}$~\cite{Garcia-Cely:2017oco}. Besides, IceCube gives stringer constraint on the lifetime of Majoron DM in the mass range $10^4~\mathrm{GeV}<M_\chi<10^9~\mathrm{GeV}$, $\Gamma_{\chi \rightarrow \nu \nu}^{-1}> \mathcal{O}\left(10^{28}\right) \sec$~\cite{IceCube:2018tkk}. So from Eq.~\eqref{decaywidth} the Majoron mass has to be smaller than $10^8~\mathrm{GeV}$ for $v_s \leq M_{\mathrm{Pl}}$. In the case without the Higgs portal, the Majoron DM can be produced by heavy neutrino decay, however, it is found to be too small to satisfy the DM relic density~\cite{Arina:2019tib,Abe:2020dut,Abe:2020ldj}. The Majoron also decays into  quarks (mainly into top quarks) through the loop diagram which we neglect in this work.

In this work, we discuss a general class of pNGB DM, rather than focusing solely on Majoron DM. Our study encompasses light DM at the MeV scale as well as superheavy DM with masses exceeding $100~\mathrm{TeV}$. The light DM scenarios naturally extend to the production of Majoron DM.

Such DM can also be produced via the freeze-in mechanism through the Higgs portal. However, in this work, we propose a novel production mechanism based on Hawking radiation and superradiance from PBHs. Since black hole particle production depends only on gravitational interactions, the pNGB DM can be produced even with feeble portal coupling. The production of pNGB DM can be realized without introducing new interactions.

\section{Schwarzschild Black Holes}\label{SchDM}
\subsection{Hawking radiation}
PBHs may originate from inhomogeneities of density perturbations in the early Universe, with these perturbations potentially arising from cosmic inflation or other alternatives. In this work, we do not focus on the formation mechanisms of PBHs but instead discuss their impact on the Universe after their formation. Assuming the temperature of the Universe at the time of PBH formation is $T=T_{\mathrm{in}}$, the initial mass of the PBH can be expressed in terms of the total radiation energy within a Hubble volume as below~\cite{Carr:2009jm}
\begin{equation}
M_{\mathrm{in}} \equiv M_{\mathrm{PBH}}\left(T_{\mathrm{in}}\right)=\frac{4 \pi}{3} \gamma \frac{\rho_r\left(T_{\mathrm{in}}\right)}{H^3\left(T_{\mathrm{in}}\right)}\,\,,
\end{equation}
where the efficiency factor is $\gamma \simeq (1/\sqrt{3})^3 \simeq 0.2$~\cite{Carr:1975qj} at the radiation epoch . Additionally, $\rho_r=\pi^2 g_* T^4 / 30$ is the radiation energy density of the Universe and $H(T)$ is the Hubble expansion rate,
\begin{equation}\label{hubble}
H(T)=\sqrt{\frac{8 \pi}{3} \frac{\rho_r}{M_{\mathrm{Pl}}^2}}=\sqrt{\frac{4 \pi^3 g_{*}}{45}} \frac{T^2}{M_{\mathrm{Pl}}}\,\,.
\end{equation}
Then we have
\begin{equation}
T_{\mathrm{in}}=\frac{1}{2}\left(\frac{5}{g_* \pi^3}\right)^{1 / 4}\left(\frac{3 \gamma M_{\mathrm{Pl}}^3}{M_{\mathrm{in}}}\right)^{1 / 2}\,\,,
\end{equation}
and from which we get $M_{\mathrm{in}} = \frac{\gamma M_{\mathrm{Pl}}^2}{2H(T_{\mathrm{in}})}$.
From Planck data combined with latest BICEP/Keck data, the tensor-to-scalar
ratio satisfies $r < 0.036$，namely, the upper limit of Hubble scale during inflation
$H(T_{\mathrm{in}})\leq 5\times10^{13}~\mathrm{GeV}$~\cite{Planck:2018jri,BICEP:2021xfz}. Then, we can obtain the lower limit of PBH initial mass $M_{\mathrm{in}} \geq 0.5~\mathrm{g}$. 

\tc{For simplicity, we assume a monochromatic mass spectrum for PBHs, such that all PBHs share the same mass upon formation. For more detailed analyses of the cosmological implications of extended mass and spin distributions, see Ref.~\cite{Cheek:2022mmy} and references therein.}
In general, we typically express the initial energy density of black holes, $\rho_{\mathrm{PBH}}\left(T_{\mathrm{in}}\right)$, as the ratio of this density to the radiation energy density. This is quantified by introducing the parameter $\beta$,
\begin{equation}\label{beta}
\beta \equiv \frac{\rho_{\mathrm{PBH}}\left(T_{\mathrm{in}}\right)}{\rho_r\left(T_{\mathrm{in}}\right)}=\frac{n_{\mathrm{PBH}}\left(T_{\mathrm{in}}\right) M_{\mathrm{in}}}{\rho_r\left(T_{\mathrm{in}}\right)}\,\,,
\end{equation}
with $n_{\mathrm{PBH}}\left(T_{\mathrm{in}}\right)$ being the initial number density of PBHs.
The properties of Schwarzschild black holes can be simply described by using only their mass. The relationship between the surface temperature of a Schwarzschild black hole, $T_{\mathrm{PBH}}$, and its mass is given by:
\begin{equation}
T_{\mathrm{PBH}}=\frac{M_{\mathrm{Pl}}^2}{8 \pi M_{\mathrm{PBH}}} \simeq 10^{13} ~\mathrm{GeV}\left(\frac{1 \mathrm{~g}}{M_{\mathrm{PBH}}}\right)\,\,.
\end{equation}
Black holes can emit particles through the process of Hawking radiation.
For an emitted species $i$ with mass $M_i$, spin $s_i$ and degrees of freedom $g_i$, its energy spectrum can be written as~\cite{Cheek:2021odj}
\begin{equation}
\frac{\mathrm{d}^2 N_i}{\mathrm{d} p \mathrm{d} t}=\frac{g_i}{2 \pi^2} \frac{\sigma_{s_i}\left(M_{\mathrm{PBH}}, M_i, p\right)}{\exp \left[E_i(p) / T_{\mathrm{PBH}}\right]-(-1)^{2 s_i}} \frac{p^3}{E_i(p)}\,\,,
\end{equation}
where $E_i(p)=\sqrt{p^2+M_i^2}$ denotes the particle energy and $\sigma_{s_i}$ is the absorption cross section which relates to the greybody factor as $\Gamma_{s_i}\equiv \sigma_{s_i}p^2/\pi$. 
It can be seen that particles are efficiently produced via Hawking radiation only when their mass is smaller than the surface temperature of the black hole. 
After summing over all particle species in the given model, and integrating over the phase space, we can get the evaporation rate of PBH mass, 
\begin{equation}\label{dMdt}
\frac{\mathrm{d} M_{\mathrm{PBH}}}{\mathrm{~d} t} =-\sum_i \int_0^{\infty} E_i \frac{\mathrm{~d}^2 N_i}{\mathrm{d} p \mathrm{d} t} \mathrm{d} p=-\varepsilon\left(M_{\mathrm{PBH}}\right) \frac{M_{\mathrm{P l}}^4}{M_{\mathrm{PBH}}^2}\,\,,
\end{equation}
where the evaporation function of PBH is defined as $\varepsilon\left(M_{\mathrm{PBH}}\right) \equiv \sum_i \varepsilon_i\left(z_i\right)$ with $z_i=M_i / T_{\mathrm{PBH}}$. It is interesting to take the approximation $T_{\mathrm{PBH}} \gg M_i$  and geometric limit $\sigma_{s_i} \rightarrow \left(27 \pi M_{\mathrm{PBH}}^2/M_{\mathrm{Pl}}^2\right)$~\cite{Page:1976df,MacGibbon:1990zk}, from which we can integrate obtain $\varepsilon_i$ analytically, 
\begin{equation}\label{eps}
\varepsilon_i\left(z_i\right)=\frac{27 g_i}{8192 \pi^5} \int_{z_i}^{\infty} \frac{x\psi_{s_i}\left(x, z_i\right)\left(x^2-z_i^2\right)}{\exp (x)-(-1)^{2 s_i}} \mathrm{~d} x \rightarrow \varepsilon_i(0)=\frac{27 g_i}{8192 \pi^5} \times \begin{cases}\frac{\pi^4}{15}, \text { Bosons } \\ \frac{7 \pi^4}{120}, \text { Fermions }\end{cases}
\end{equation}
where $x=E_i / T_{\mathrm{PBH}}$, and $\psi_{s_i} \equiv \sigma_{s_i} /\left(27 \pi M_{\mathrm{PBH}}^2/M_{\mathrm{Pl}}^2\right)$. 
So the mass loss rate of a PBH is approximately,
\begin{equation}
\frac{d M_{\mathrm{PBH}}}{d t}=-\varepsilon\left(M_{\mathrm{PBH}}\right) \frac{M_{\mathrm{Pl}}^4}{M_{\mathrm{PBH}}^2} \approx-\frac{27}{4} \frac{g_{* \mathrm{PBH}}}{30720 \pi} \frac{M_{\mathrm{Pl}}^4}{M_{\mathrm{PBH}}^2}\,\,,
\end{equation}
which is consistent with \cite{Baldes:2020nuv}. $g_{* \mathrm{PBH}}$ is the total degree of freedom emitted by the PBH.
From this equation we can get
\begin{equation}
M_{\mathrm{PBH}}(t)=M_{\mathrm{in}}\left(1-\frac{t-t_{\mathrm{in}}}{\tau}\right)^{1 / 3}\,\,,
\end{equation}
with $\tau=\frac{40960 \pi}{27 g_{* \mathrm{PBH}}} \frac{M_{\mathrm{in}}^3}{M_{\mathrm{Pl}}^4}$ being the PBH lifetime.

Assuming the Universe is always radiation-dominated, utilizing the relationship between the Hubble constant and temperature, along with $H(\tau)= 1 / (2 \tau)$, the Universe's temperature $T_{\mathrm{evap}}$ when PBH evaporates can be determined,
\begin{equation}\label{Tevap}
T_{\mathrm{evap}} \simeq \frac{9}{128}\left(\frac{1}{5 g_* \pi^5}\right)^{1 / 4}\left(\frac{g_{* \mathrm{PBH}} M_{\mathrm{Pl}}^5}{2 M_{\mathrm{in}}^3}\right)^{1 / 2}\,\,.
\end{equation}
During the radiation-dominated era, the energy density of black holes evolves as $T^3$, while the radiation energy density evolves as $T^4$. Consequently, when $\beta$ exceeds a critical value $\beta_c \equiv T_{\text{evap}}/T_{\text{in}}$, PBHs will dominate the Universe before their evaporation.

In the case that PBHs dominate the Universe's energy density before they completely evaporate, $T_{\text {evap}}$ is determined by setting $H(\tau)=2 /(3 \tau)$ and making the replacement $\rho_r \rightarrow M_{\mathrm{in}} n_{\mathrm{PBH}}\left(T_{\mathrm{evap}}\right) = M_{\mathrm{in}}n_{\text{PBH}}\left(T_{\text{in}}\right) T_{\text{evap}}^3 / T_{\text{in}}^3$ in Eq.~\eqref{hubble}, then we get
\begin{equation}
\left.T_{\mathrm{evap}}\right|_{\mathrm{PBHdom}} \simeq \frac{9}{256}\left(\frac{g_{* \mathrm{PBH}}^2}{2 \beta}\right)^{1 / 3}\left(\frac{1}{5^5 \pi^{17} g_*^3}\right)^{1 / 12}\left(\frac{M_{\mathrm{Pl}}^{17}}{3 \gamma M_{\mathrm{in}}^{11}}\right)^{1 / 6}\,\,.
\end{equation}
To be consistent with the observational results of BBN, we require the initial mass of PBHs to be less than $\sim 10^9~\mathrm{g}$ such that they fully evaporate before BBN. Furthermore, if PBHs dominate the Universe prior to their evaporation, they may generate a significant amount of induced GWs, contributing additional radiation energy density and thus impacting BBN observations. The constraint on the initial fraction is given by~\cite{Domenech:2020ssp}
\begin{equation}\label{betaGW}
    \beta \lesssim 1.1 \times 10^{-6} \left(\frac{\gamma}{0.2}\right)^{-1/2} \left(\frac{M_{\mathrm{in}}}{10^4~\mathrm{g}}\right)^{-17/24}\,\,.
\end{equation}
In the massless limit and geometric limit, the production rate of particle species $i$ is
\begin{equation}
\Gamma_{\mathrm{PBH}\rightarrow i} \equiv \frac{d N_i}{d t}=\frac{27 g_i M_{\mathrm{Pl}}^2}{1024 \pi^4 M_{\mathrm{PBH}}} \begin{cases}2 \zeta(3), & \text { Bosons } \\ \frac{3}{2} \zeta(3), & \text { Fermions }\end{cases}\,\,.
\end{equation}
For any particle $i$, by using $\int d t=\int d M_{\mathrm{PBH}}\left(d M_{\mathrm{PBH}} / d t\right)^{-1}$, if the initial PBH temperature $T_{\mathrm{in}}$ is larger than the mass $M_i$ where we assume $i$ is a boson,
\begin{equation}
\begin{aligned}
	N_i  =\int_0^{M_{\mathrm{in}}} \frac{1}{\varepsilon\left(M_{\mathrm{PBH}}\right)} \frac{M_{\mathrm{PBH}}^2}{M_{\mathrm{Pl}}^4} \frac{27 g_i \zeta(3) M_{\mathrm{Pl}}^2}{512 \pi^4 M_{\mathrm{PBH}}} d M_{\mathrm{PBH}} 
	 =\frac{120 \zeta(3)}{\pi^3} \frac{g_i}{g_{* \mathrm{PBH}}}\left(\frac{M_{\mathrm{in}}}{M_{\mathrm{Pl}}}\right)^2\,\,.
\end{aligned}
\end{equation}
While, if $T_{\mathrm{in}}<M_i$, the PBH can radiate $i$ only when the PBH temperature raises to exceed $M_i$,
\begin{equation}
\begin{aligned}
	N_i  =\int_0^{\frac{M_{\mathrm{Pl}}^2}{8 \pi M_{i}}} \frac{1}{\varepsilon\left(M_{\mathrm{PBH}}\right)} \frac{M_{\mathrm{PBH}}^2}{M_{\mathrm{Pl}}^4} \frac{27 g_i \zeta(3) M_{\mathrm{Pl}}^2}{512 \pi^4 M_{\mathrm{PBH}}} d M_{\mathrm{PBH}} 
	 =\frac{15 \zeta(3)}{8 \pi^5} \frac{g_i}{g_{* \mathrm{PBH}}}\left(\frac{M_{\mathrm{Pl}}}{M_i}\right)^2\,\,,
\end{aligned}
\end{equation}
if $i$ is a fermion, the above expressions have to be multiplied by $3 / 4$. 

If the PBHs dominate the energy density of the Universe before evaporation, the radiation into SM particles will reheat the Universe and dilute the final DM relic.
The dilution factor can be derived by using the conservation of energy before and after the evaporation of PBHs~\cite{Bernal:2022pue},
\begin{equation}
	\frac{\pi^2}{30} g_{*} T_{\text {evap}}^4+M_{\text{in}} \frac{n_{\mathrm{PBH}}\left(T_{\text {in}}\right)}{s\left(T_{\text{in}}\right)} s\left(T_{\text{evap}}\right)  =\frac{\pi^2}{30} g_{*} \tilde{T}_{\text {evap}}^4 \,\,,
\end{equation}
from which we get
\begin{equation}
    \frac{s(\tilde{T}_{\text {evap}})}{s\left(T_{\text {evap }}\right)}=\left(\frac{\tilde{T}_{\text{evap}}}{T_{\text {evap }}}\right)^3=\left(1+\frac{\beta T_{\text {in }}}{T_{\text {evap }}}\right)^{3 / 4}\,\,,
\end{equation}
where $s(T)$ is the entropy density and $\tilde{T}_{\mathrm {evap}}$ is the temperature of the SM plasma after PBH evaporation occurs. The PBHs come to dominate the Universe's energy density before they evaporate if the second factor in the bracket exceeds unity. Then the particle number density $Y_i\equiv n_i(T)/s(T)$ produced from PBHs is diluted and reads,
\begin{equation}
Y_i \equiv \frac{N_i n_{\mathrm{PBH}}\left(T_{\text{evap}}\right)}{s\left(T_{\text{evap}}\right)}\frac{s\left(T_{\mathrm{evap}}\right)}{s(\tilde{T}_{\mathrm{evap}})}=\frac{N_i n_{\mathrm{PBH}}\left(T_{\text{in}}\right)}{s\left(T_{\text{in}}\right)}\frac{s\left(T_{\mathrm{evap}}\right)}{s(\tilde{T}_{\mathrm{evap}})}=\frac{3 \beta T_{\mathrm{in}} N_i}{4 M_{\mathrm{in}}} \frac{s\left(T_{\mathrm{evap}}\right)}{s(\tilde{T}_{\mathrm{evap}})}\,\,.
\end{equation}
The relic density of the particle $i$ at present is
\begin{equation}
\Omega_{i}^{\mathrm{evap}}=\frac{\rho_{i}}{\rho_c}=\frac{M_{i}}{\rho_c} \frac{n_{i}}{s}\left(t_0\right) s_0=\frac{M_{i}}{\rho_c} Y_{i} s_0\,\,,
\end{equation}
where $\rho_c \simeq 4.78 \times 10^{-6}~\mathrm{GeV}~\mathrm{cm}^{-3}$ and $
s_0\simeq 2891.2~\mathrm{cm}^{-3}$~\cite{Planck:2018vyg} are the present critical energy density and entropy density, respectively.
In summary, if there is no PBH dominated period, namely, $\beta<\beta_c$, the relic density can be obtained as
\begin{align}
	\Omega_{i}^{\mathrm{evap}} & \simeq 1.64 \times 10^6 \beta g_i \left(\frac{g_*}{100}\right)^{-1 / 4}\left(\frac{g_{* \mathrm{PBH}}}{100}\right)^{-1 }\left(\frac{M_{\mathrm{in}}}{M_{\mathrm{Pl}}}\right)^{1 / 2}\left(\frac{M_{i}}{\mathrm{GeV}}\right), \quad M_{i}<\frac{M_{\mathrm{Pl}}^2}{8 \pi M_{\mathrm{in}}}\,\,, \label{omega1}\\
	& \simeq 2.59 \times 10^3\beta g_i\left(\frac{g_*}{100}\right)^{-1 / 4}\left(\frac{g_{* \mathrm{PBH}}}{100}\right)^{-1 }\left(\frac{M_{\mathrm{Pl}}^7}{M_{\mathrm{in}}^3 M_{i}^4}\right)^{1 / 2}\left(\frac{M_{i}}{\mathrm{GeV}}\right), \quad M_{i}>\frac{M_{\mathrm{Pl}}^2}{8 \pi M_{\mathrm{in}}}\,\,.\label{omega2}
\end{align}
While, if the initial energy fraction of PBH is large enough, namely, $\beta>\beta_c$, the corresponding relic density reads,
\begin{align}
	\Omega_{i}^{\mathrm{evap}} & \simeq 6.12 \times 10^5g_i\left(\frac{g_*}{100}\right)^{-1 / 4}\left(\frac{g_{* \mathrm{PBH}}}{100}\right)^{-1 / 2}\left(\frac{M_{\mathrm{Pl}}}{M_{\mathrm{in}}}\right)^{1 / 2}\left(\frac{M_{i}}{\mathrm{GeV}}\right), \quad M_{i}<\frac{M_{\mathrm{Pl}}^2}{8 \pi M_{\mathrm{in}}}\,\,, \label{omega3}\\
	& \simeq 0.97 \times 10^3g_i\left(\frac{g_*}{100}\right)^{-1 / 4}\left(\frac{g_{* \mathrm{PBH}}}{100}\right)^{-1 / 2}\left(\frac{M_{\mathrm{Pl}}^9}{M_{\mathrm{in}}^5 M_{i}^4}\right)^{1 / 2}\left(\frac{M_{i}}{\mathrm{GeV}}\right), \quad M_{i}>\frac{M_{\mathrm{Pl}}^2}{8 \pi M_{\mathrm{in}}}\,\,.\label{omega4}
\end{align}
It is interesting to notice that when PBHs dominate the energy density of the early Universe before they evaporate, the relic density is independent of $\beta$.
\subsection{UV freeze-in}
In addition to being produced via Hawking evaporation, we note that in the model, when $\lambda_{HS}$ is nonzero, DM can also be generated through the freeze-in mechanism via the Higgs portal $HH \rightarrow s \rightarrow \chi + \chi$. The decay width of the scalar field $s$ is~\cite{Abe:2020ldj},
\begin{equation}
    \Gamma_{s \rightarrow H^\dagger H} = \frac{\lambda_{HS}^2 M_s}{8\pi \lambda_S} \sqrt{1-\frac{4M_h^2}{M_s^2}}, \quad \Gamma_{s \rightarrow \chi \chi} = \frac{\lambda_{S} M_s}{32\pi } \sqrt{1-\frac{4M_\chi^2}{M_s^2}}\,\,.
\end{equation}
with $M_h \simeq 125~\mathrm{GeV}$ being the mass of SM Higgs. During freeze-in, the Boltzmann equation for the pNGB DM is given by~\cite{Abe:2020dut,Abe:2020ldj}
\begin{equation}
	\frac{d Y_\chi^{\mathrm{FI}}(x)}{d z}=\frac{2}{s(z) H(z) z} \gamma_{\chi \chi \rightarrow H^{\dagger} H}\,\,,
\end{equation}
where $z=M_\chi/T$ and $\gamma_{\chi \chi \rightarrow H^{\dagger} H}$ is the interaction rate,
\begin{equation}
	\gamma_{\chi \chi \rightarrow H^{\dagger} H} \equiv \frac{g_H}{2!2!} \frac{T \lambda_{H S}^2}{2^9 \pi^5} \int_{4 M_\chi^2}^{\infty} d s \sqrt{s-4 M_\chi^2} K_1(\sqrt{s} / T) \frac{s^2}{\left(s-M_s^2\right)^2} \simeq \frac{3\lambda_{HS}^2 T^8}{2\pi^5 M_s^4}\,\,.
\end{equation}
with $g_H$ being the degree of freedom of SM Higgs.
Assuming the DM begins to be produced at plasma temperature $T_{\mathrm{reh}}$, in the UV Freeze-in scheme, $M_s \gg T_{\mathrm{reh}}$,
\begin{equation}
	\Omega^{\mathrm{FI}}_\chi  = 3.34 \times 10^{22} \lambda_{H S}^2 \left(\frac{100}{g_s}\right) \left(\frac{100}{g_*}\right)^{1/2} \left(\frac{T_{\mathrm{reh}}}{M_s} \right)^4 \left( \frac{M_\chi}{T_{\mathrm{reh}}}\right)\,\,.
\end{equation}
We choose $\lambda_{HS} = 10^{-7}$, $M_s = 10~T_{\mathrm{reh}}$ and $T_{\mathrm{reh}} = T_{\mathrm{in}}$ when considering UV freeze-in in this work. We also have to multiply it with the dilution factor  $\left(1+\beta \frac{T_{\mathrm{in}}}{T_{\text {evap }}}\right)^{-3 / 4}$ when we take into account the reheating from PBHs if the PBHs dominate the Universe before they evaporate. When $M_s \ll T_{\mathrm{in}}$, the pNGB DM could also be produced by IR freeze-in and the relic reads~\cite{Abe:2020ldj},
\begin{equation}
    Y_\chi^{\mathrm{IR}} \approx \frac{405 \sqrt{10}}{(2 \pi)^5} \frac{\mathrm{Br}_{s \rightarrow \chi \chi} \lambda_{H S}^2 M_{\mathrm{Pl}}}{g_{*s} g_*^{1 / 2} \lambda_{S} M_s} \sqrt{1-\frac{4 M_h^2}{M_s^2}} \,\,.
\end{equation}
where $\mathrm{Br}_{s \rightarrow \chi \chi} = \Gamma_{s \rightarrow \chi \chi} / (\Gamma_{s \rightarrow H^\dagger H}+\Gamma_{s \rightarrow \chi \chi} )$ is the branch ratio of the decay of $s$ into the DM.
It can be seen that the result does not depend on the initial temperature $T_{\mathrm{in}}$, so this contribution can dominate over the PBH evaporation if we fine-tune the coupling $\lambda_{HS}$. However, in this work we focus on the DM production from PBH so we only consider the UV freeze-in where $M_s>T_{\mathrm{in}}$.
It is worth noting that PBHs can not only directly produce DM but can also generate intermediate mediators $s$, which subsequently decay into two DM particles.
We require $\lambda_S \leq 10^{-6}$ and $v_s\leq M_{\mathrm{Pl}}$ where the former condition comes from the non-equilibrium of DM.  So the scalar mass should satisfy $M_s\leq 10^{-3}M_{\mathrm{Pl}}$. From this and $M_s = 10~T_{\mathrm{in}}$ we get $M_{\mathrm{in}} \gtrsim 3~\mathrm{g}$.

In the early Universe, the DM also suffers the contribution from gravitational freeze-in where the gravitons play the role of mediator. The DM is produced after reheating where we assumed $T_\mathrm{reh} = T_\mathrm{in}$ then the Boltamann equation is
\begin{equation}
	\frac{d Y_\chi^{\mathrm{GR}}(x)}{d z}=\frac{2}{s(z) H(z) z} \gamma_{g} \,\,,
\end{equation}
where $\gamma_{g} =64\pi^2 \delta \frac{ T^8}{M_{\mathrm{Pl}}^4}$ with $\delta  = \frac{3997 \pi^3}{41472000}$ for boson DM~\cite{Clery:2021bwz}.
The DM yield then reads 
\cite{Bernal:2018qlk,Tang:2017hvq,Garny:2017kha,Clery:2021bwz},
\begin{equation}
    Y_\chi^{\mathrm{GR}} = \frac{720 \delta}{\pi^2 g_{*s}}\sqrt{\frac{5\pi}{g_*}}\left(\frac{T_{\mathrm{in}}}{M_{\mathrm{Pl}}}\right)^3 \frac{s\left(T_{\mathrm{evap}}\right)}{s(\tilde{T}_{\mathrm{evap}})}\times \begin{cases}1, \quad M_\chi \ll T_{\mathrm{in}} \\ \left(\frac{T_{\mathrm{in}}}{M_\chi}\right)^4, \quad M_\chi \gg T_{\mathrm{in}}\end{cases}
\end{equation}
Finally the relic density of DM from gravtational production is $\Omega_\chi^{\mathrm{GR}} = Y_\chi^{\mathrm{GR}}M_\chi s_0 /\rho_c$.

Thus, in general, the final DM relic density is
\begin{equation}
    \Omega_{\mathrm{DM}} = \Omega^{\mathrm{evap}}_\chi + 2\mathrm{Br}_{s \rightarrow \chi \chi} \Omega^{\mathrm{evap}}_s + \Omega^{\mathrm{FI}}_\chi+\Omega^{\mathrm{GR}}_\chi\,\,.
\end{equation}
In the following, we will use this formula to consider two different scenarios for DM production, namely, the light pNGB DM and heavy pNGB DM. The present DM relic abundance satisfies $\Omega_{\mathrm{DM}}h^2 \simeq 0.12$ where the dimensionless Hubble parameter $h \approx 0.67$.

\begin{figure}[h]
        \centering
	\begin{minipage}{0.8\linewidth}
		\vspace{3pt}
		\centerline{\includegraphics[width=0.9\linewidth]{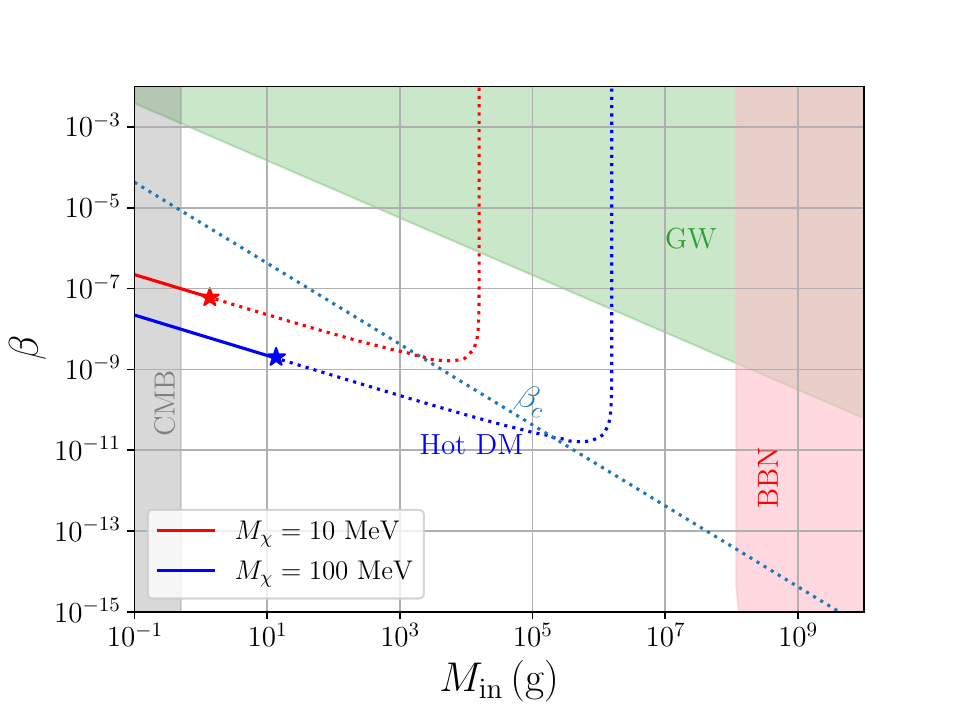}}
	\end{minipage}
	\caption{Contours showing the parameter space where PBHs produce the correct relic abundance of DM. The red and blue curves correspond to DM masses of $10~\mathrm{MeV}$ and $100~\mathrm{MeV}$, respectively. The star marks the critical black hole mass above which the evaporated DM particles are hot DM. Gray, green, and red regions represent excluded regions from CMB observations, excessive GW, and BBN, respectively.}
	\label{fig:Slight}
\end{figure}

\subsection{Light pNGB dark matter}
Firstly, we consider the case of light DM with mass $M_\chi = 10~\mathrm{MeV}$ and $100~\mathrm{MeV}$. The DM relic density in these cases is given by Eqs.~\eqref{omega1} and \eqref{omega3}. We find that the UV and gravitational freeze-in contribution are negligible. 

For light DM, radiated DM may be too hot. 
We must carefully examine its velocity to determine whether it is excluded by large-scale structure observation.
We can simply follow the discussions on the constraints of warm DM.
The average DM velocity today $V_0$ is given by~\cite{Fujita:2014hha}
 \begin{equation}\label{vpre}
     V_0 = \frac{a_{\mathrm{evap}}}{a_0} \frac{\langle p_\chi^{\mathrm{evap}} \rangle}{M_\chi} \simeq \frac{a_{\mathrm{evap}}}{a_{\mathrm{eq}}} \frac{a_{\mathrm{eq}}}{a_0}\frac{\langle E_\chi^{\mathrm{evap}} \rangle}{M_\chi}\,\,,
 \end{equation}
 where $\langle p_\chi^{\mathrm{evap}}\rangle$ and $\langle E_\chi^{\mathrm{evap}}\rangle$ are the averaged momentum and averaged energy of DM at evaporation, respectively. $a_{\mathrm{evap}}$ is the scale factor after PBH evaporation. $a_0$ and $a_{\mathrm{eq}}$ are the scale factor at present and  matter-radiation equality, respectively. Note that we have the relation
 \begin{equation}\label{redshift}
     \frac{a_{\mathrm{evap}}}{a_{\mathrm{eq}}} \simeq  \frac{T_{\mathrm{eq}}}{\tilde{T}_{\mathrm{evap}}}\left[\frac{g_{* s}\left(T_{\mathrm{eq}}\right)}{g_{* s}\left(\tilde{T}_{\mathrm{evap}}\right)}\right]^{\frac{1}{3}}, \quad \frac{a_{\mathrm{eq}}}{a_0} \simeq \frac{\Omega_{r}}{\Omega_{m}} \approx 3\times 10^{-4}\,\,.
 \end{equation}
Then the main task is to calculate the averaged energy of pNGB DM which is produced by PBH evaporation.
\begin{equation}
    \langle E_\chi^{\mathrm{evap}} \rangle = \frac{1}{N_\chi}\int E_\chi \frac{d N_\chi}{\mathrm{d}p\mathrm{d}t} \mathrm{d}p\mathrm{d}t\simeq \frac{\pi^3}{120 \zeta(3)}\frac{g_{*\mathrm{PBH}}}{g_\chi}\frac{M_{\mathrm{Pl}}^2}{M_{\mathrm{in}}^2} \int  \frac{d M_{\mathrm{PBH}}}{dt} \frac{g_\chi}{g_{*\mathrm{PBH}}} \mathrm{d}t\approx 5.4 \left(\frac{M_{\mathrm{Pl}}^2}{8\pi M_{\mathrm{in}}}\right)\,\,.
\end{equation}
In the second equality we have used Eq.~\eqref{dMdt} and the fact that all particles including DM are lighter than the Hawking temperature of PBHs. The value of the prefactor is close to 6 given by~\cite{Morrison:2018xla} and 6.3 by \cite{Baldes:2020nuv}. By using $V_0 \lesssim 1.8\times 10^{-8}$ from the upper limits of the warm DM velocity~\cite{Masina:2020xhk},
\begin{equation}
    M_\chi > 7.3\times 10^{-3}~\mathrm{GeV} \times \left(\frac{M_{\mathrm{in}}}{1~\mathrm{g}}\right)\,\,.
\end{equation}
\tc{So for a specific value of DM mass, the PBH mass is bounded from above to satisfy the large-scale structure. This is because from Eqs.~\eqref{Tevap} and \eqref{redshift} we have $a_{\mathrm{evap}}/a_{\mathrm{eq}}\propto T_{\mathrm{eq}}/T_{\mathrm{evap}}\propto M_{\mathrm{in}}^{3/2}$ which means that lighter PBH evaporates earlier. Smaller value of PBH mass corresponds to smaller value of $a_{\mathrm{evap}}/a_{\mathrm{eq}}$ and then larger redshift. Although lighter PBHs possess higher Hawking temperatures, the average momentum of the emitted DM is sufficiently redshifted by cosmic expansion, rendering the DM cold at present.}
The results are shown in Fig.~\ref{fig:Slight}, where the star denotes the critical black hole mass. For black holes exceeding the critical mass, the produced DM becomes too hot to satisfy the constraint from large-scale structure formation.
The gray region is from the CMB constraint and the green region is from Eq.~\eqref{betaGW}.
The pink region represents the BBN constraint which is taken from the most recent result \cite{Boccia:2024nly}.

\begin{figure}[h]
	\begin{minipage}{0.52\linewidth}
		\vspace{3pt}
		\centerline{\includegraphics[width=\textwidth]{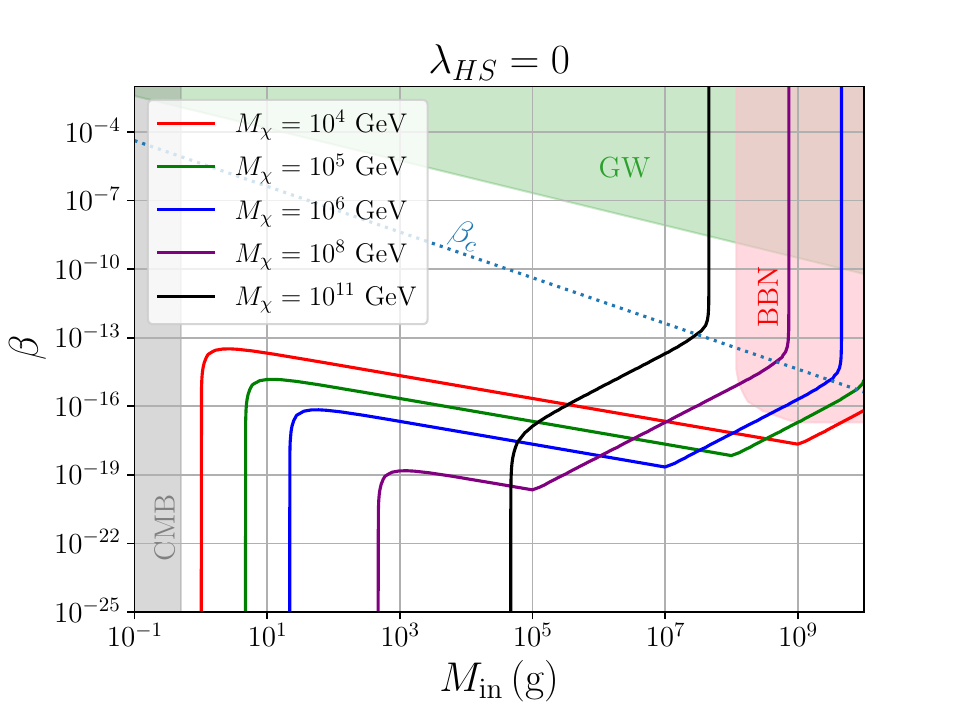}}
	\end{minipage}
	\begin{minipage}{0.52\linewidth}
		\vspace{3pt}
		\centerline{\includegraphics[width=\textwidth]{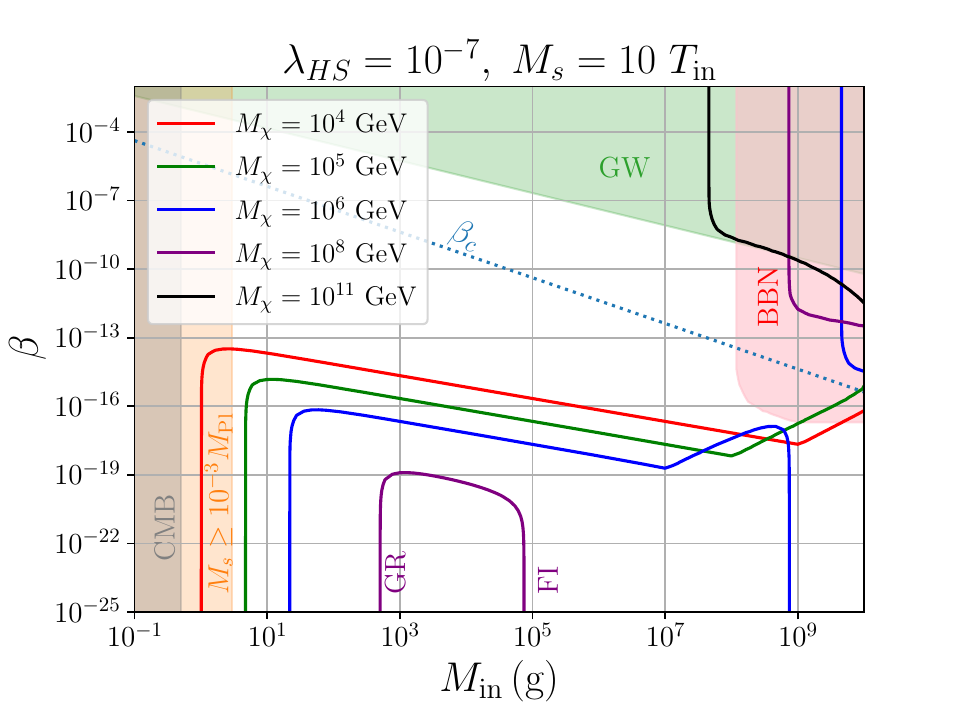}}
	\end{minipage}
	\caption{Contour plot for heavy pNGB DM. The left panel shows the scenario where DM is produced from the evaporation of Schwarzschild PBHs and gravitational production, without considering the contribution from UV freeze-in. The right panel includes the contribution from UV freeze-in. ``GR" and ``FI" represent the regions where the DM relic are dominated by gravitational and UV freeze-in, respectively. For the UV freeze-in process, we choose $\lambda_{HS} = 10^{-7}$ and $M_s = 10~T_{\mathrm{in}}$.}
	\label{fig:Sheavydm}
\end{figure}

\subsection{Heavy pNGB dark matter}
For the heavy pNGB DM case, we show the main results in Fig.~\ref{fig:Sheavydm}.
In the left panel, we show the results without UV freeze-in.
The red, green, blue, purple, and black lines correspond to pNGB masses of $10^4$, $10^5$, $10^6$, $10^8$, and $10^{11}~\mathrm{GeV}$, respectively. The gravitational production dominates when the PBH initial mass is small enough. The decrease of $\beta$ with $M_{\mathrm{in}}$ is described by Eq.~\eqref{omega1}, where $\beta \propto (M_{\mathrm{in}}/\mathrm{GeV})^{-1/2}$. And then, when the initial PBH mass is larger, the DM relic density satisfies Eq.~\eqref{omega1} where $\beta \propto (M_{\mathrm{in}}/\mathrm{GeV})^{3/2}$. When $\beta > \beta_c$ the PBHs dominate the Universe and the relic density is independent of $\beta$. In the right panel, we include the UV freeze-in with $\lambda_{HS} = 10^{-7}$. The orange region does not satisfy the condition that $M_s \leq 10^{-3}M_{\mathrm{Pl}}$ for $\lambda_S \leq 10^{-6}$ and $v_s \leq M_{\mathrm{Pl}}$. The UV freeze-in dominates when the $M_{\mathrm{in}}$ is larger, or in other words, when the $T_{\mathrm{in}}$ is lower.

\section{Kerr Black Holes}\label{KerDM}
\subsection{Hawking radiation and superradiance}

Kerr black holes, due to their non-zero angular momentum, alter their surface temperature and radiation rate and also the spectrum of Hawking radiation. This deviation from the static Schwarzschild case provides a new perspective for understanding quantum effects in strong gravitational fields.
For the case of Kerr black holes, the horizon temperature reads,
\begin{equation}
T_{\mathrm{PBH}}=\frac{M_{\mathrm{Pl}}^2}{4 \pi M_{\mathrm{PBH}}} f(a_\star)=\frac{M_{\mathrm{Pl}}^2}{4 \pi M_{\mathrm{PBH}}} \frac{\sqrt{1-a_{\star}}}{1+\sqrt{1-a_{\star}}}\,\,,
\end{equation}
where $a_{\star} \equiv J M_{\mathrm{Pl}}^2 / M_{\mathrm{PBH}}^2$, with $J$ denoting the angular momentum of PBHs. The case with $a_{\star}=0$ corresponds to the Schwarzschild limit. The energy spectrum now explicitly depends on the angular quantum numbers $l$ and $m$~\cite{Cheek:2021odj},
\begin{equation}\label{dNdtKerr}
\frac{\mathrm{d}^2 N_i}{\mathrm{~d} p \mathrm{d} t}=\frac{g_i}{2 \pi^2} \sum_{l=s_i} \sum_{m=-l}^l \frac{\sigma_{s_i}^{l m}\left(M_{\mathrm{PBH}}, p, a_{\star}\right)}{\exp \left[\left(E_i(p)-m \Omega\right) / T_{\mathrm{PBH}}\right]-(-1)^{2 s_i}} \frac{p^3}{E_i(p)}\,\,,
\end{equation}
with $\Omega=\left(a_{\star}M_{\mathrm{Pl}}^2 /\left(2 M_{\mathrm{PBH}}\right)\right)\left(1 /\left(1+\sqrt{1-a_{\star}^2}\right)\right)$ being the horizon angular velocity.

For Kerr black holes, due to the NGB nature, pNGB DM can be produced not only via Hawking radiation but also through black hole superradiance.
The superradiance of DM happens when the Compton length of the DM particle $\lambda_C = 1/M_\chi$ is comparable with the gravitational radius $r_g = M_{\mathrm{PBH}}/M_{\mathrm{Pl}}^2$. The superradiance process is efficient when \tc{$\alpha \equiv M_{\mathrm{PBH}}M_\chi/M_{\mathrm{Pl}}^2 =r_g/\lambda_C \sim \mathcal{O}(1)$}. 
Considering the dominant unstable mode with quantum number $n=2$ and $l=m=1$, the superradiance rate $\Gamma_{\mathrm{SR}}$ can be approximated by the Detweiler formalism~\cite{Detweiler:1980uk,Dolan:2007mj},
\begin{equation}\label{det}
\Gamma_{\mathrm{SR}}=\frac{M_{\chi}}{24}\left(\frac{M_{\mathrm{PBH}} M_{\chi}}{M_{\mathrm{Pl}}^2}\right)^8\left(a_{\star}-2 M_{\chi} r_{+}\right)\,\,,
\end{equation}
where $r_{+}=M_{\mathrm{PBH}}\left(1+\sqrt{1-a_\star^2}\right)/M_{\mathrm{Pl}}^2$ represents the event horizon. It is clear that $\Gamma_{\text {SR }}$ vanishes in the limit $a_{\star} \rightarrow 0, r_{+} \rightarrow 0$. This approximation \tc{is only valid in the  non-relativistic limit and} fails in the large $\alpha$ regime, where we need to solve the Klein-Gordon equation on the Kerr background numerically~\cite{Dolan:2007mj}. \tc{However, the analytic expression in Eq.~\eqref{det} agrees with the numerical results to within a factor of a few, rendering it sufficient for the purposes of qualitative analysis in this work.}
Therefore, for the PBH with $M_{\mathrm{PBH}} \lesssim 10^{9}~\mathrm{g}$, only heavy pNGB with $M_\chi \gtrsim 10^4~\mathrm{GeV}$
will get contribution from superradiance.

The final amount of DM particles $N_{\chi}^{\mathrm{SR}}(\mathrm{end})$ per PBH after the superradiance is easy to obtain~\cite{Bernal:2022oha},
\begin{equation}
N_{\chi}^{\mathrm{SR}}(\mathrm{final}) \simeq 1.5 \times 10^{19}\left(\frac{M_{\mathrm{in}}}{10^5 ~\mathrm{g}}\right)^2\left(a_{\star}^{\mathrm{in}}-a_{\star}^{\mathrm{end}}\right) \simeq \mathcal{O}\left(10^{18}\right)\left(\frac{M_{\mathrm{in}}}{10^5 ~\mathrm{g}}\right)^2\,\,.
\end{equation}
The typical timescale for superradiance
\begin{equation}
t_{\mathrm{SR}} \sim \frac{1}{\Gamma_{\mathrm{SR}}} \simeq \frac{24}{M_{\chi}}\left(\frac{M_{\mathrm{Pl}}^2}{M_{\mathrm{in}} M_{\chi}}\right)^8 \frac{1}{a_{\star}^{\mathrm{in}}}\,\,.
\end{equation}
For the DM production from Kerr black hole, we need to perform numerical calculations. 
In the following, we define
\begin{equation}
\mathcal{N}_\chi = n_\chi a^3, \quad \varrho_{\mathrm{PBH}} = \rho_{\mathrm{PBH}} a^3, \quad \varrho_{r} = \rho_{r} a^4\,\,.
\end{equation}
which can help to simplify the Boltzmann equations.

\subsection{Light pNGB dark matter}
Unlike the Schwarzschild black holes, for Kerr black holes we cannot derive approximately analytic results. Instead, we have to solve the evolution equations of PBHs numerically.
For light DM, since superradiance is ineffective in this case, we only need to consider black hole evolution and DM production driven by Hawking radiation. From Eq.~\eqref{dNdtKerr} we get,
\begin{align}
Ha \frac{d M_{\mathrm{PBH}}}{d a} & =-\varepsilon(M_{\mathrm{PBH}}, a_\star) \frac{M_{\mathrm{Pl}}^4}{M_{\mathrm{PBH}}^2} \label{light1}\\
Ha \frac{d a_{\star}}{d a} & =-a_{\star}[\gamma(M_{\mathrm{PBH}},a_\star)-2 \varepsilon(M_{\mathrm{PBH}},a_\star)] \frac{M_{\mathrm{Pl}}^4}{M_{\mathrm{PBH}}^3}\,\,, \label{light2}\\
Ha \frac{d\mathcal{N}_\chi^{\mathrm{PBH}}}{da} & =\frac{\varrho_{\mathrm{PBH}}}{M_{\mathrm{PBH}}}\Gamma_{\mathrm{PBH} \rightarrow \chi}\,\,,\label{light3}
\end{align}
where $\varepsilon\left(M_{\mathrm{PBH}}, a_{\star}\right)$ and $\gamma\left(M_{\mathrm{BH}},a_\star\right)$ are given by
\begin{equation}
\begin{aligned}
	& \varepsilon\left(M_{\mathrm{PBH}}, a_{\star}\right)\equiv \sum_i\varepsilon_i\left(z_i, a_{\star}\right)=\frac{27}{8192 \pi^5} \sum_i\int_{z_i}^{\infty} \sum_{l, m} \frac{\psi_{s_i}^{l m}\left(x, a_{\star}\right)\left(x^2-z_i^2\right)}{\exp \left(x^{\prime} / 2 f\left(a_{\star}\right)\right)-(-1)^{2 s_i}} x \mathrm{~d} x \\
	& \gamma\left(M_{\mathrm{PBH}},a_\star\right) \equiv \sum_i \gamma_i\left(z_i, a_{\star}\right)=\frac{27}{1024 \pi^4} \sum_i\int_{z_i}^{\infty} \sum_{l, m} \frac{m \psi_{s_i}^{l m}\left(x, a_{\star}\right)\left(x^2-z_i^2\right)}{\exp \left(x^{\prime} / 2 f\left(a_{\star}\right)\right)-(-1)^{2 s_i}} \mathrm{~d} x\,\,,
\end{aligned}
\end{equation}
with $x^{\prime}=x-m \Omega^{\prime}$, where $x=8 \pi M_{\mathrm{PBH}} E_i/M_{\mathrm{Pl}}^2$, $\Omega^{\prime}=\Omega / T_{\mathrm{PBH}}$. 

It is also required to write down the evolution equations of the radiation energy of the Universe and the total energy of PBHs as
\begin{align}
\frac{d \varrho_{r}}{d a} & =-\frac{\varepsilon_{\mathrm{SM}}(M_{\mathrm{PBH}},a_\star)}{\varepsilon(M_{\mathrm{PBH}},a_\star)} \frac{1}{M_{\mathrm{PBH}}} \frac{d M_{\mathrm{PBH}}}{d a} a \varrho_{\mathrm{PBH}}\,\,, \label{energy1}\\
\frac{d \varrho_{\mathrm{PBH}}}{d a} & =\frac{1}{M_{\mathrm{PBH}}} \frac{d M_{\mathrm{PBH}}}{d a} \varrho_{\mathrm{PBH}}, \label{energy2}\\
H^2 & =\frac{8 \pi }{3M_{\mathrm{Pl}}^2}\left(\varrho_{\mathrm{PBH}} a^{-3}+\varrho_{r} a^{-4}\right)\,\,. \label{energy3}
\end{align}
where $\varepsilon_{\mathrm{SM}}$ only includes the contributions from SM particles.
The source terms on the right-hand side of the first and second lines are provided by PBH Hawking radiation and superradiance. The third line represents the Hubble rate of cosmic expansion.
Besides，we also have to track the evolution of the temperature of the Universe~\cite{Lunardini:2019zob,Bernal:2019lpc,Arias:2019uol},
\begin{equation}\label{tem}
\frac{d T}{d a}=-\frac{T}{\Delta}\left\{\frac{1}{a}+\frac{\varepsilon_{\mathrm{SM}}(M_{\mathrm{PBH}},a_\star)}{\varepsilon(M_{\mathrm{PBH}},a_\star)} \frac{1}{M_{\mathrm{PBH}}} \frac{d M_{\mathrm{PBH}}}{d a} \frac{g_*(T)}{g_{* s}(T)} \frac{a \varrho_{\mathrm{PBH}}}{4 \varrho_{R}}\right\}\,\,,
\end{equation}
where
\begin{equation}\label{var}
\Delta=1+\frac{T}{3 g_{*s}(T)} \frac{d g_{* s}(T)}{d T}
\end{equation}
takes into account the variation of the degree of freedoms.

\begin{figure}[h]
	\begin{minipage}{0.52\linewidth}
		\vspace{3pt}
		\centerline{\includegraphics[width=\textwidth]{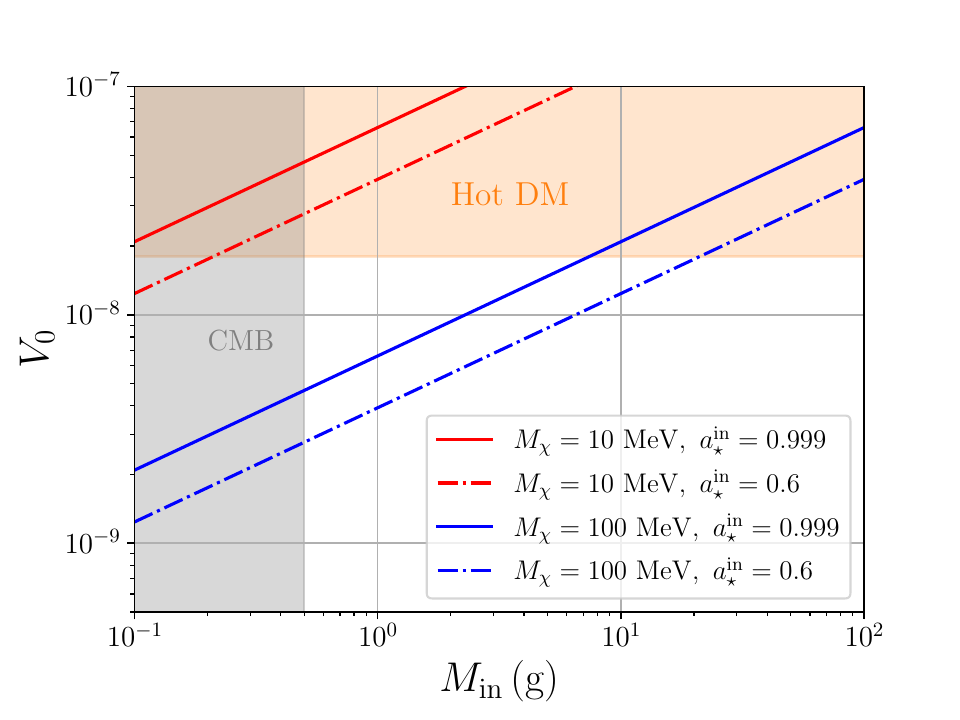}}
	\end{minipage}
	\begin{minipage}{0.52\linewidth}
		\vspace{3pt}
		\centerline{\includegraphics[width=\textwidth]{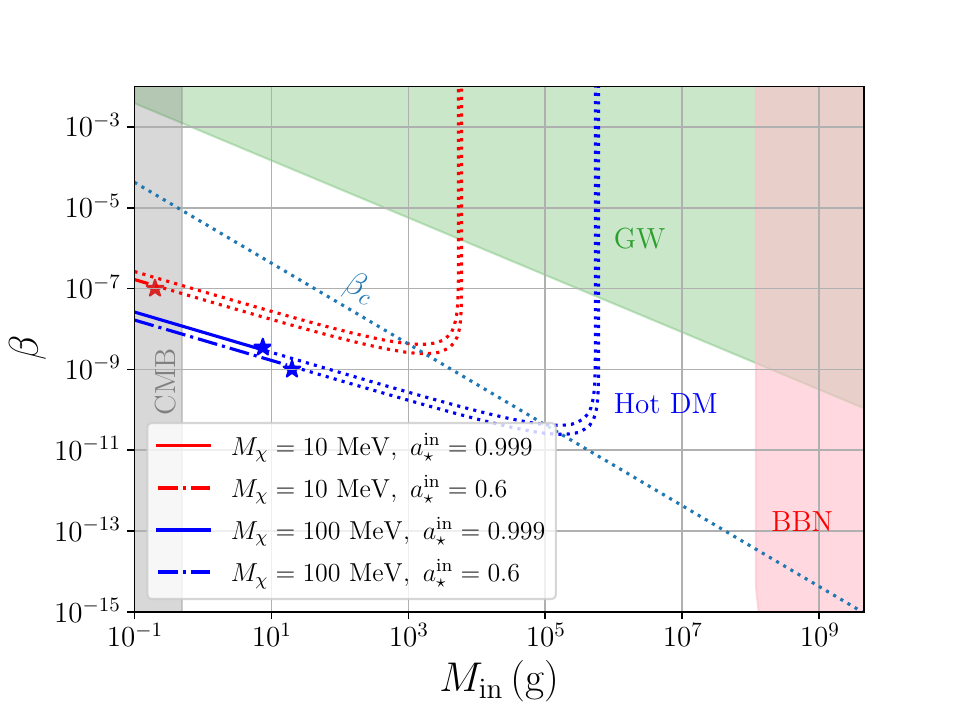}}
	\end{minipage}
	\caption{Left panel: The present averaged velocity of DM particles evaporated from PBHs. The orange region represents exclusions from Lyman-$\alpha$ observations. Right panel: Contour plot for the relic density of light pNGB DM in the case of Kerr black holes. The stars represent the critical values of PBH mass beyond which the DM is too hot to satisfy the large scale structure.}
	\label{fig:kerrlightdm}
\end{figure}

We solve Eqs.~\eqref{light1}-\eqref{light3} and Eqs.~\eqref{energy1}-\eqref{tem} to obtain the PBHs' evolution. The equations are solved numerically by using the Python package ULYSSES~\cite{Cheek:2021odj,Granelli:2020pim,Granelli:2023vcm}. Then we can get
\begin{equation}
    \langle E_\chi^{\mathrm{evap}} \rangle = \frac{1}{N_\chi}\int_{t_{\mathrm{in}}}^{t_{\mathrm{evap}}}  E_\chi \frac{\mathrm{d} N_\chi}{\mathrm{d}p \mathrm{d}t} \mathrm{d}p\mathrm{d}t\simeq \frac{\int \mathrm{d}t ~\varepsilon_\chi(M_\chi/T_{\mathrm{PBH}},a_\star)M_{\mathrm{Pl}}^4 M_{\mathrm{PBH}}^{-2}}{\int \mathrm{d}t ~\Gamma_{\mathrm{PBH}\rightarrow \chi}(M_{\mathrm{PBH}},a_\star)}\,\,,
\end{equation}
where $t_{\mathrm{evap}}$ and the evolution of PBHs are given by solving the differential equations numerically. Then we can integrate the right side to get the averaged energy, which is substituted in Eq~\eqref{vpre} to get the present velocity of DM.

In the left panel of Fig.~\ref{fig:kerrlightdm} we show the averaged velocity for different values of DM mass and initial angular momentum of PBHs, as functions of the PBH initial mass. We can see that the $10~\mathrm{MeV}$ pNGB DM is totally excluded by combined constraints from CMB and large scale structure. The averaged velocity is enhanced for larger PBH mass and larger angular momentum.
The right panel in Fig.~\ref{fig:kerrlightdm} shows the contour plot of DM relic density. It can be seen that the shapes are only slightly different from the results in Fig.~\ref{fig:Slight}, demonstrating the validity of the analytical results.

\begin{figure}[h]
	\centering
	\begin{minipage}{0.49\linewidth}
		\centering
		\includegraphics[width=0.9\linewidth]{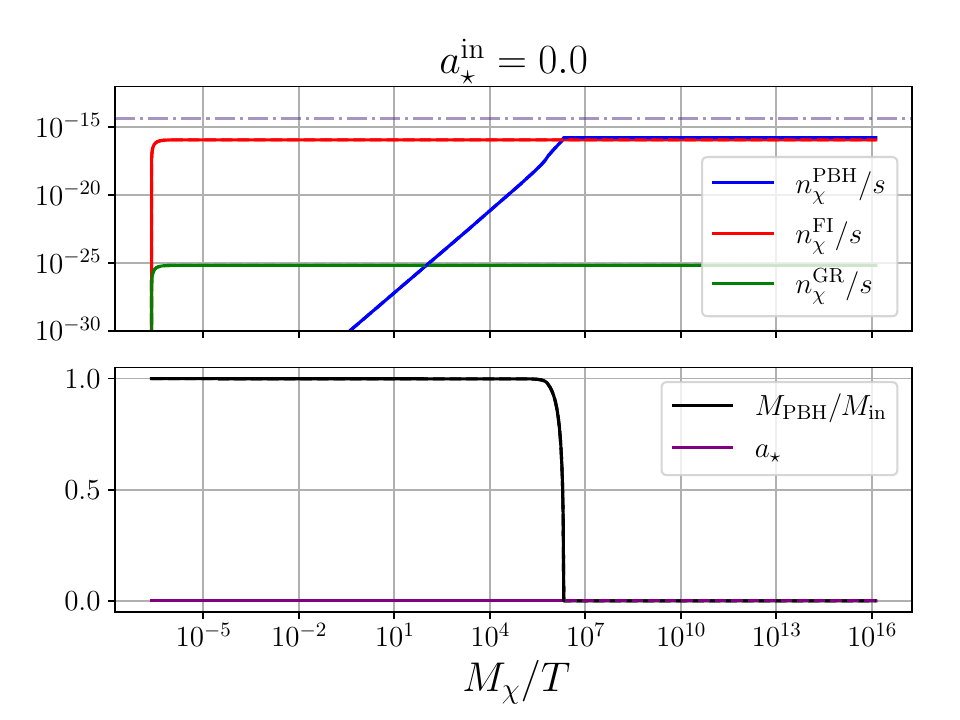}
	\end{minipage}
	\begin{minipage}{0.49\linewidth}
		\centering
		\includegraphics[width=0.9\linewidth]{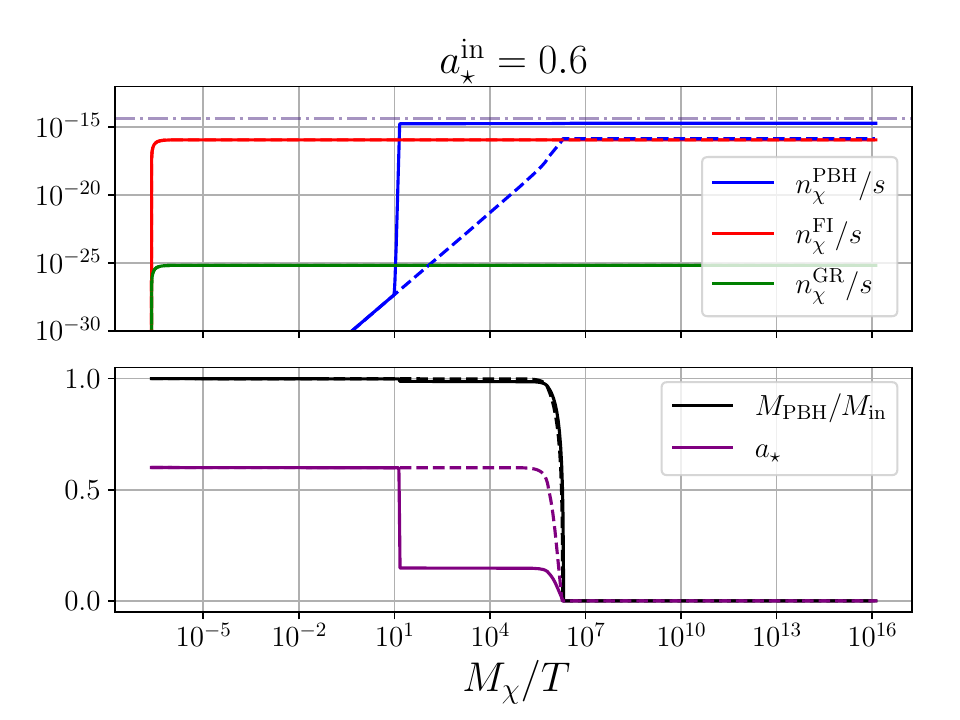}
	\end{minipage}
	\qquad
	
	\begin{minipage}{0.49\linewidth}
		\centering
		\includegraphics[width=0.9\linewidth]{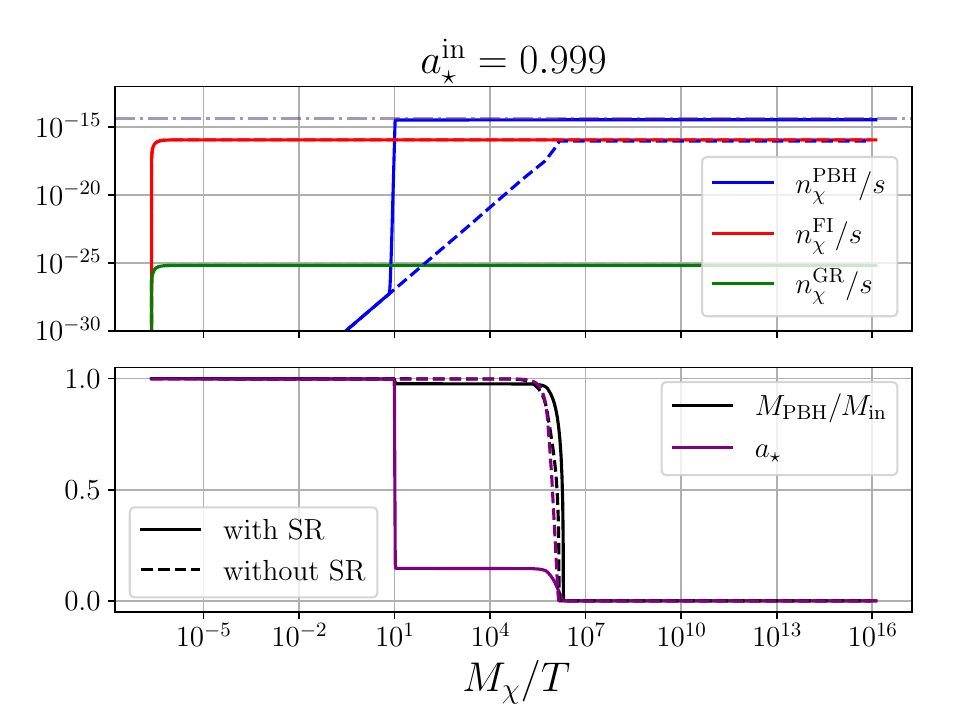}
		\label{chutian3}
	\end{minipage}
        \caption{Evolution of DM relic $n_\chi/s(T)$, PBH's mass and PBH's angular momentum as functions of $M_\chi/T$. The blue lines and the red lines represent the DM from PBH and UV freeze-in, respectively. The green lines represent the contribution from gravitational production. The black lines and the purple lines represent the evolution of PBH's mass and angular momentum, respectively. The solid lines describe the case taking the superradiance into account and the dashed lines describe the case without superradiance. The gray horizontal lines correspond to the correct DM relic abundance. We set $M_{\mathrm{in}} = 10^8~\mathrm{g}$, $\beta = 10^{-20}$ and $M_\chi = 10^5~\mathrm{GeV}$.}
        \label{fig:relicdensity}
\end{figure}

\subsection{Heavy pNGB dark matter}

Now we consider the production of heavy DM. The main difference is that, in this case, the pNGB DM is produced not only by the Hawking evaporation but also the superradiance \tc{when the Compton length of DM is comparable with the gravitational radius. We will see that when $\alpha \in (0.1,1)$, the DM is dominantly produced by superradiance.}. It has been shown in Ref.~\cite{Bernal:2022oha} that the superradiant production enhances the DM relic density by a factor of $\mathcal{O} (10)$. The superradiant production has also been used for the leptogenesis mechanism~\cite{Ghoshal:2023fno} and the production of dark radiation~\cite{Manno:2025dhw}.

In the case of pNGB DM produced by Hawking radiation and superradiance, the Boltzmann equations for the PBH-DM system read,
\begin{align}
Ha \frac{d N_{\chi}^{\mathrm{SR}}}{d a} & =\Gamma_{\mathrm{SR}} N_{\chi}^{\mathrm{SR}} \label{NSR}\,\,,\\
Ha \frac{d M_{\mathrm{PBH}}}{d a} & =-\varepsilon(M_{\mathrm{PBH}},a_\star) \frac{M_{\mathrm{Pl}}^4}{M_{\mathrm{PBH}}^2}-M_{\chi} \Gamma_{\mathrm{SR}} N_{\chi}^{\mathrm{SR}} \,\,,\\
Ha \frac{d a_{\star}}{d a} & =-a_{\star}[\gamma(M_{\mathrm{PBH}},a_\star)-2 \varepsilon(M_{\mathrm{PBH}},a_\star)] \frac{M_{\mathrm{Pl}}^4}{M_{\mathrm{PBH}}^3} \notag\\
&\quad -\left[\sqrt{2}-2 \alpha a_{\star}\right] \Gamma_{\mathrm{SR}} N_{\chi}^{\mathrm{SR}} \frac{M_{\mathrm{Pl}}^2}{M_{\mathrm{PBH}}^2}\,\,, \\
Ha \frac{d\mathcal{N}_\chi^{\mathrm{PBH}}}{da} & =\frac{\varrho_{\mathrm{PBH}}}{M_{\mathrm{PBH}}}\left[\Gamma_{\mathrm{PBH} \rightarrow \chi}+\Gamma_{\mathrm{SR}} N_{\chi}^{\mathrm{SR}}\right] \,\,,
\end{align}
and the equations for DM produced by UV and gravitational freeze-in are,
\begin{align}\label{FI}
Ha \frac{d\mathcal{N}_\chi^{\mathrm{FI}}}{da} = 2a^3 \gamma_{\chi \chi \rightarrow H^{\dagger} H}, \quad Ha \frac{d\mathcal{N}_\chi^{\mathrm{GR}}}{da} = 2a^3 \gamma_{g}\,\,.
\end{align}
$N_\chi$ represents the total number of pNGB DM particles produced via superradiance by an individual black hole. The first line describes the number of pNGB DM particles generated through black hole superradiance. The second and third lines illustrate the decay of black hole mass and angular momentum, accounting for both Hawking radiation and superradiance. The fourth formula gives the total number of DM particles produced through Hawking radiation and superradiance. Finally, the UV freeze-in mechanism for generating pNGB DM is included.
We solve Eqs.~\eqref{energy1}-\eqref{tem} and Eqs.~\eqref{NSR}-\eqref{FI} numerically to obtain the DM relic abundance and the evolution of PBHs.

\begin{figure}[h]
	\begin{minipage}{0.52\linewidth}
		\vspace{3pt}
		\centerline{\includegraphics[width=\textwidth]{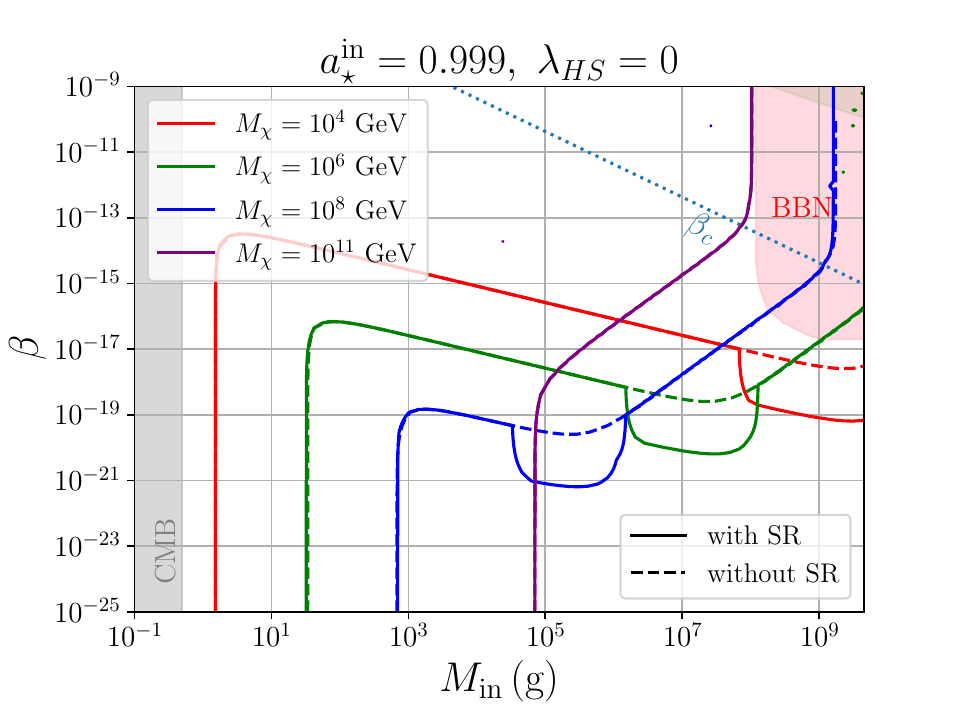}}
	\end{minipage}
	\begin{minipage}{0.52\linewidth}
		\vspace{3pt}
		\centerline{\includegraphics[width=\textwidth]{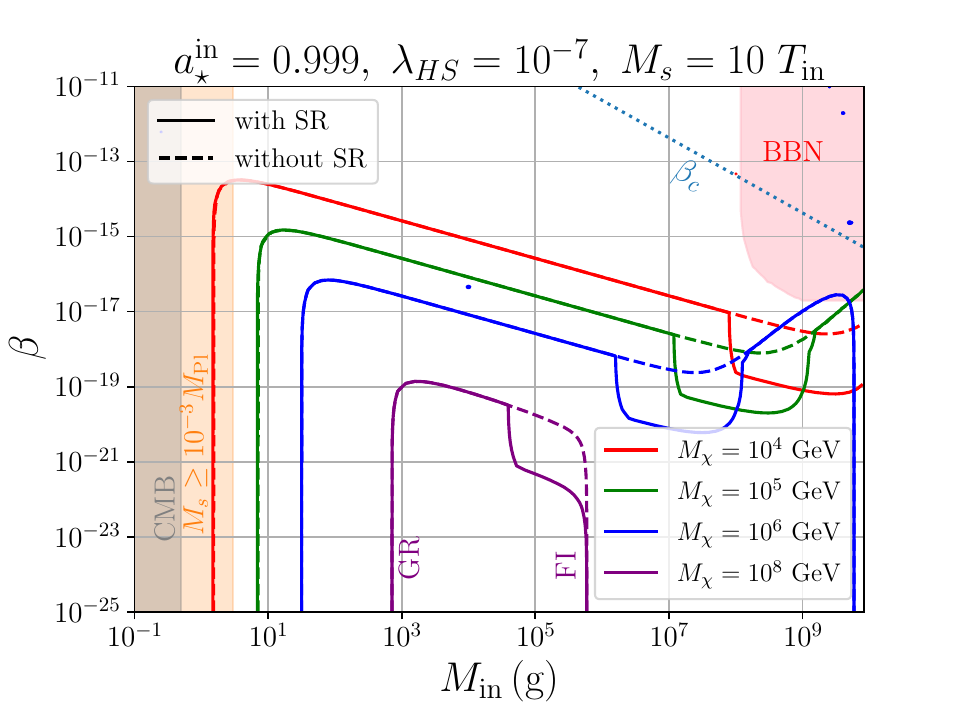}}
	\end{minipage}
	\caption{Left panel: Contour plot showing regions where the relic abundance of DM satisfies correct DM relic density $\Omega_{\mathrm{DM}}h^2 = 0.12$, considering contributions from Hawking radiation, gravitational production, and superradiance. The ``SR'' represents the superradiance. The dashed lines represent the cases without superradiance. Right panel: Contour plot showing regions where the relic abundance of DM satisfies $\Omega_{\mathrm{DM}}h^2 = 0.12$, accounting for Hawking radiation, superradiance, UV freeze-in and gravitational production.}
	\label{fig:kerrheavydm}
\end{figure}

As an example,  We set the parameters $M_{\mathrm{in}} = 10^8~\mathrm{g}$, $\beta = 10^{-20}$, $M_\chi = 10^5~\mathrm{GeV}$ and the numerical results are shown in Fig.~\ref{fig:relicdensity}. We show the evolution of DM relic density from PBHs and freeze-in as functions of $M_\chi/T$. And we also show the evolution of the mass and angular momentum of PBHs. The solid and dashed lines represent the cases with and without superradiance, respectively. Then we scan over the parameter space and show the contour plot of correct relic density of DM for different values of DM mass in Fig.~\ref{fig:kerrheavydm}. The initial angular momentum parameter of PBH is fixed to be $a_\star^{\mathrm{in}} = 0.999$. Compared to the left panel, the right panel includes UV freeze-in with $\lambda_{HS} = 10^{-7}$. It can be seen that the superradiance plays an important role at $\alpha \sim \mathcal{O}(0.1)$. For Kerr black holes with $a_{\star}^{\mathrm{in}} = 0.999$, the superradiance enhances the DM relic abundance by a factor of $\mathcal{O}(10-100)$.

\section{Gravitational wave}\label{GWs}
\subsection{Gravitational wave from domain wall}

The explicit symmetry-breaking term in the Lagrangian reduces the symmetry of the model to a $Z_2$ symmetry. Consequently, the breaking of $Z_2$ symmetry leads to the formation of DWs in the Universe. However, since DWs are inherently stable, their energy density would eventually dominate the Universe, which is clearly not physically acceptable. It is known that quantum gravity does not respect any global symmetries. To address this issue, we may introduce the following term \cite{Rai:1992xw,Borah:2024kfn}.
\begin{equation}
    \Delta V = \frac{1}{\Lambda_{\mathrm{QG}}}\left(c_1 S^5 + c_2 S^3 H^2 +c_3 S H^4 \right)\,\,,
\end{equation}
where $\Lambda_{\mathrm{QG}}$ denotes the scale of QG and the prefactors $c_{1,2,3}$ are some constants. Because we have $v_s \gg v$, the bias term is dominated by the first term and we can absorbed the prefactors into $\Lambda_{QG}$, then  one can write the energy bias after
 replacing fields with their VEVs
\begin{equation}
  \Delta V \simeq \frac{1}{\Lambda_{\mathrm{QG}}} v_s^5\,\,.
\end{equation}
Require the DWs annihilate away before they dominate the Universe, we get the following condition~\cite{Saikawa:2017hiv}:
\begin{equation}
    \Delta V^{1/4} > 2.18\times 10^{4}~\mathrm{GeV} \times C_{\mathrm{ann}}^{1/4} \mathcal{A}^{1/2} \left(\frac{\sigma}{10^{27}~\mathrm{GeV}^3}\right)^{1/2}\,\,,
\end{equation}
where $\sigma = \frac{4}{3} \sqrt{\lambda_S} v_s^3$ is the surface tension of the DW, $C_{\mathrm{ann}} \simeq 2$ is a dimensionless constant, and $\mathcal{A} \simeq 0.8 \pm 0.1$ is the area parameter. In terms of the breaking scale, this condition translates to
\begin{equation}
    v_s < 1.95\times 10^{12}~\mathrm{GeV} \left(\frac{10^{30}~\mathrm{GeV}}{\Lambda_{\mathrm{QG}}} \right) \left(\frac{10^{-6}}{\lambda_S} \right)\,\,.
\end{equation}
Besides, the DWs could decay into SM particles and destroy the predictions of BBN. In order to avoid this, the is another constraint on the bias energy~\cite{Saikawa:2017hiv}：
\begin{equation}
    \Delta V^{1/4} > 16.03~\mathrm{GeV} \times C_{\mathrm{ann}}^{1/4} \mathcal{A}^{1/4} \left(\frac{\sigma}{10^{27}~\mathrm{GeV}^3}\right)^{1/4}\,\,,
\end{equation}
or equivalently,
\begin{equation}
    v_s > 375.32~\mathrm{GeV} \left(\frac{\lambda_S}{10^{-6}}\right)^{1/4}\left(\frac{\Lambda_{\mathrm{QG}}}{10^{30}~\mathrm{GeV}}\right)\,\,,
\end{equation}
under which the DWs can annihilate away before the BBN. After the DWs annihilate, the peak frequency of the resulting GW is given by~\cite{Saikawa:2017hiv,Chen:2020wvu,Bhattacharya:2023kws}
\begin{equation}
\begin{aligned}
    &f_p^{\mathrm{DW}} \simeq \\
    &3.75~\times 10^{-4}~\mathrm{Hz}~C_{\mathrm{ann}}^{-1/2} \mathcal{A}^{-1/2} \left(\frac{g_*\left(T_{\mathrm{ann}}\right)}{10}\right)^{1 / 4}\left(\frac{g_{* s}\left(T_{\mathrm{ann}}\right)}{10}\right)^{-1 / 3}\left(\frac{10^{27}~\mathrm{GeV}^3}{\sigma}\right)^{1/2}\left( \frac{\Delta V}{10^{16}~\mathrm{GeV}^4}\right)^{1/2}\,\,,\\
    &\simeq 2.57~\mathrm{Hz} \left(\frac{g_*\left(T_{\mathrm{ann}}\right)}{10}\right)^{1 / 4}\left(\frac{g_{* s}\left(T_{\mathrm{ann}}\right)}{10}\right)^{-1 / 3}\left(\frac{\lambda_S}{10^{-6}}\right)^{-1/4}\left(\frac{v_s}{10^{12}~\mathrm{GeV}}\right)\left(\frac{\Lambda_{\mathrm{QG}}}{10^{30}~\mathrm{GeV}}\right)^{-1/2}\,\,,
\end{aligned}
\end{equation}
and peak energy density spectrum of DW collapse,
\begin{equation}
    \Omega_p^{\mathrm{DW}} h^2 \simeq 5.3\times 10^{-4} \times \epsilon_{\mathrm{GW}} C_{\mathrm{ann}}^{2}\mathcal{A}^{4}\left(\frac{g_{* s}\left(T_{\mathrm{ann}}\right)}{10}\right)^{-4 / 3}\left(\frac{\sigma}{10^{27}~\mathrm{GeV}^3}\right)^4 \left(\frac{10^{16}~\mathrm{GeV}^4}{\Delta V}\right)^2\,\,,
\end{equation}
where $\epsilon_{\mathrm{GW}} \simeq 0.7$~\cite{Hiramatsu:2013qaa} represents the  fraction of energy radiated into GW. We adopt the following power-law spectrum for the DW GW signals~\cite{Caprini:2019egz,NANOGrav:2023hvm},
\begin{equation}
    \Omega_{\mathrm{GW}}^{\mathrm{DW}} h^2 = \Omega_p^{\mathrm{DW}} h^2 \frac{(a+b)^c}{(bx^{-a/c}+ax^{b/c})^c}\,\,,
\end{equation}
where $x\equiv f/f_p^{\mathrm{DW}}$ and $a=3$. Numerical simulations suggest that $b \simeq c\simeq 1$~\cite{Hiramatsu:2013qaa}. \tc{Note that the DW GW spectrum often takes the assumption that the network collapse happens almost instantly. However, more precise GW spectrum should be investigated by using numerical simulations of biased DWs~\cite{Cyr:2025nzf,Notari:2025kqq,Babichev:2025stm}.}

\begin{figure}[h]
	\begin{minipage}{0.52\linewidth}
		\vspace{3pt}
		\centerline{\includegraphics[width=\textwidth]{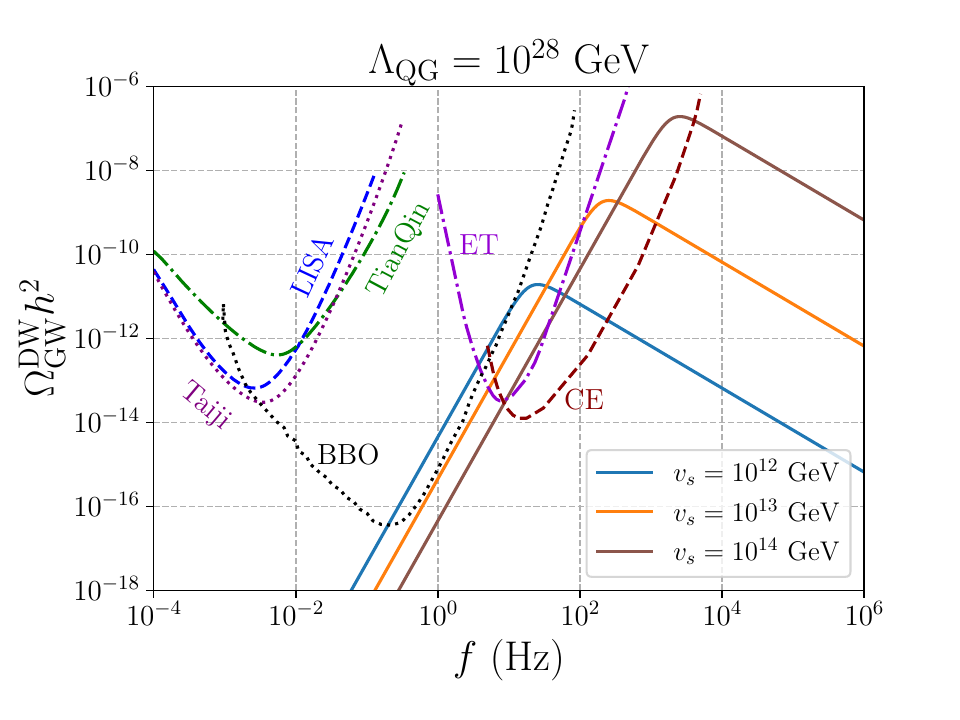}}
	\end{minipage}
        \begin{minipage}{0.52\linewidth}
		\vspace{3pt}
		\centerline{\includegraphics[width=\textwidth]{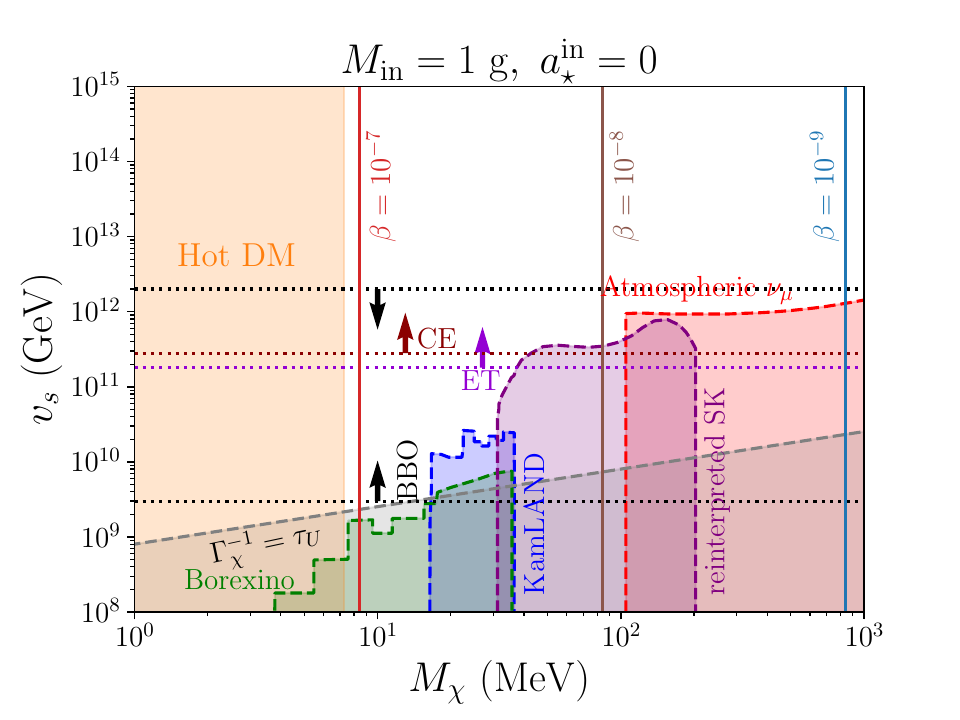}}
	\end{minipage}
	\caption{Left panel: The DW GW spectrum for $v_s = 10^{12}$, $10^{13}$ and $10^{14}~\mathrm{GeV}$ and the sensitivity
 curves of LISA, Tianqin, Taiji, BBO, CE and ET. Right panel: The case of $M_{\mathrm{in}} = 1~\mathrm{g}$ for Schwarzschild black holes. The region between the two black lines is accessible to BBO experiments assuming $\mathrm{SNR}>10$, while areas above the dark red and dark purple lines are detectable by CE and ET, respectively. The green, blue, purple and red regions are excluded by Borexino, KamLAND, reinterpreted SK, and atmospheric $\nu_\mu$ spectra. On the vertical lines the correct DM abundance is fulfilled with the corresponding values of $\beta$.}
	\label{fig:dwgw}
\end{figure}

It has been argued that the successful formation of DWs requires the mass of real scalar $M_s$ has to be smaller than the reheating temperature $T_{\mathrm{in}}$. So the case with PBH and UV freeze-in can not produce detectable DW GWs. However, the DW GWs can detect the pNGB DM produced purely from PBHs instead. For given GW spectrum $\Omega_{\mathrm{GW}}h^2$, to quantify the detectability of the GW signals, we calculate the signal-to-noise ratio~\cite{Moore:2014lga,Breitbach:2018ddu,Azatov:2019png},
\begin{equation}
    \mathrm{SNR} = \sqrt{\mathcal{T}\int_{f_{\mathrm{min}}}^{f_{\mathrm{max}}}df\left(\frac{\Omega_{\mathrm{GW}}h^2}{\Omega_{\mathrm{sens}}h^2}\right)^2}\,\,,
\end{equation}
where $\Omega_{\mathrm{sens}}h^2$ is the noise spectrum of the experiments and $\mathcal{T}$ is the duration time of the experiment. We assume the GW is detectable when $\mathrm{SNR} > 10$. In the left panel of Fig.~\ref{fig:dwgw} we fix $\Lambda_{\mathrm{QG}} = 10^{28}~\mathrm{GeV}$ and show the DW GWs for $v_s = 10^{12}$, $10^{13}$, $10^{14}~\mathrm{GeV}$. The GW signals could be detected by BBO~\cite{Crowder:2005nr,Corbin:2005ny},  Cosmic Explorer (CE)~\cite{Reitze:2019iox} and Einstein telescope 
 (ET)~\cite{Sathyaprakash:2012jk,ET:2019dnz}. If the breaking scale $v_s$ is lower, the signals can also be probed by LISA~\cite{LISA:2017pwj}, TianQin~\cite{TianQin:2015yph} and Taiji~\cite{Hu:2017mde}. 
 
 Notably, assuming the pNGB is Majoron, the DM can directly decay into the neutrino flux, which is constrained by dedicated experiments such as Borexino~\cite{Borexino:2010zht}, KamLAND~\cite{KamLAND:2011bnd}, reinterpreted SK~\cite{Palomares-Ruiz:2007egs}, and constraint from atmospheric $\nu_\mu$ spectra~\cite{Palomares-Ruiz:2007egs}. Here we assume the decay width of pNGB is given by Eq.~\eqref{decaywidth}. The results are shown in the right panel of Fig.~\ref{fig:dwgw}. We can see that due to the strong constraints from these experiments, the VEV $v_s$ associated with pNGB DM must be large enough such that the peak frequencies of the corresponding GW signals are higher than 1~Hz. Additionally, although the DW GWs can also help to detect heavy pNGB DM, here we only show the parameter space of the light MeV pNGB. Constraints on heavier Majoron DM can be found in Ref.~\cite{Akita:2023qiz}. The excluded regions from Borexino, KamLAND, reinterpreted SK and atmospheric $\nu_\mu$ spectra in the right panel of Fig.~\ref{fig:dwgw} are taken from Ref.~\cite{Garcia-Cely:2017oco}.

By the way, it should be noted that the DW annihilation itself can produce the pNGB and PBHs~\cite{Garriga:2015fdk,Deng:2016vzb,Gelmini:2023ngs,Lazarides:2023ksx} which we leave in our future work.

\subsection{Induced gravitational wave}
\begin{figure}[h]
	\begin{minipage}{0.52\linewidth}
		\vspace{3pt}
		\centerline{\includegraphics[width=\textwidth]{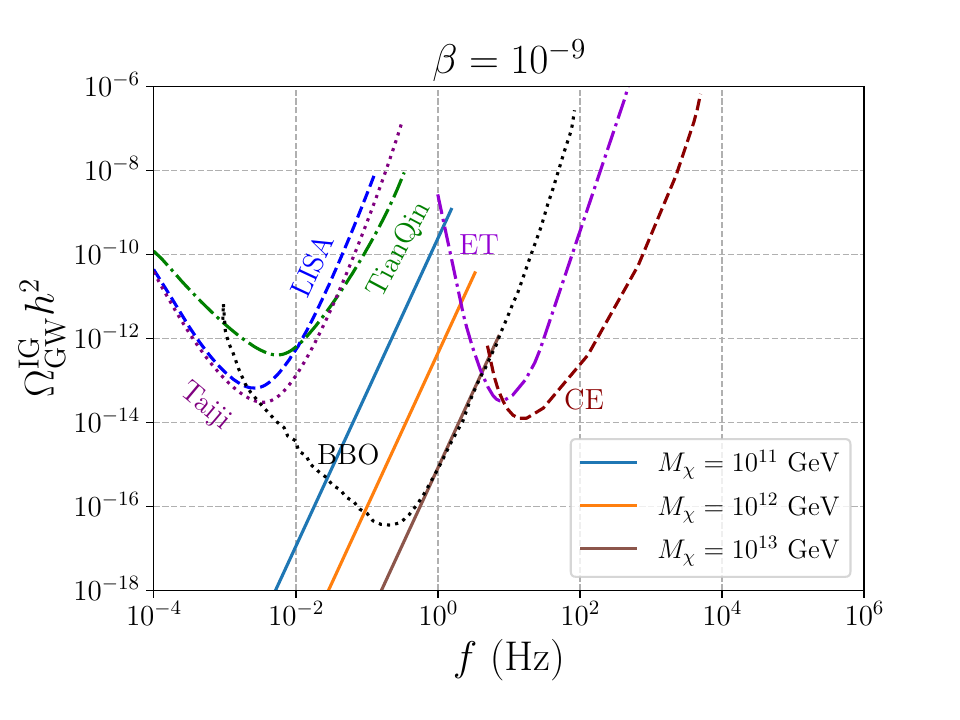}}
	\end{minipage}
        \begin{minipage}{0.52\linewidth}
		\vspace{3pt}
		\centerline{\includegraphics[width=\textwidth]{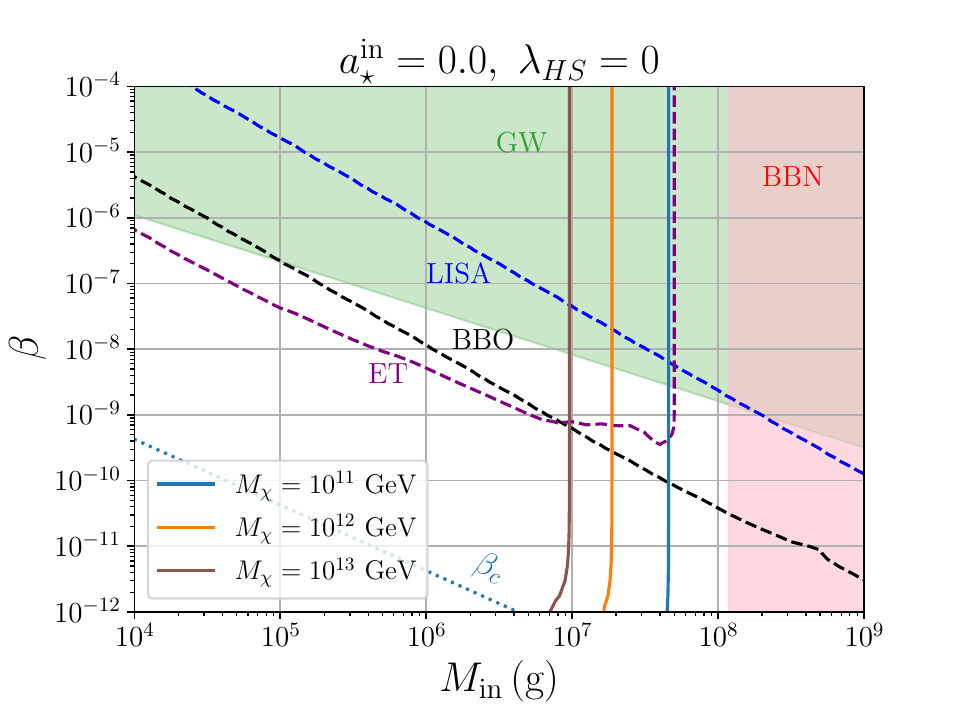}}
	\end{minipage}
	\caption{Left panel: Induced GW spectrum for $M_\chi=10^{11}$, $10^{12}$ and $10^{13}~\mathrm{GeV}$. Right panel: The contour plot of superheavy DM relic density where the DM mass is $M_\chi=10^{11}$, $10^{12}$ and $10^{13}~\mathrm{GeV}$. We take the Schwarzschild case with $a_\star^{\mathrm{in}} = 0$ and $\lambda_{HS}=0$. The regions above the black, blue, and purple dashed lines represent areas that can be observed by GW experiments with $\mathrm{SNR}>10$.}
	\label{fig:ingw}
\end{figure}

After PBHs form, they distribute randomly throughout the Universe following Poissonian statistics~\cite{Papanikolaou:2020qtd}. When the PBHs begin to dominate the energy density of the Universe, this inhomogeneous distribution of PBHs leads to curvature perturbations. Subsequently, at second order, these perturbations can generate GWs. The present-day induced GW spectrum can be written as \cite{Domenech:2020ssp,Borah:2022iym}
\begin{equation}\label{induced}
\Omega_{\mathrm{GW}}^{\mathrm{IG}} h^2\simeq \Omega_{p}^{\mathrm{IG}}h^2\left(\frac{f}{f_{{p}}^{\mathrm{IG}}}\right)^{11 / 3} \Theta\left(f_{{p}}^{\mathrm{IG}}-f\right)\,\,,
\end{equation}
where
\begin{equation}
\Omega_{p}^{\mathrm{IG}}h^2 \simeq 9 \times 10^{-7}\left(\frac{\beta}{10^{-8}}\right)^{16 / 3}\left(\frac{M_{\mathrm{in}}}{10^7~\mathrm{g}}\right)^{34 / 9}
\end{equation}
indicates the peak amplitude of GW spectrum.
The $f_{{p}}^{\mathrm{IG}}$ in Eq.~\eqref{induced} corresponds to the cutoff that is derived from the mean separation between PBHs, below which the PBHs cannot be viewed as continued fluid anymore. The peak frequency is found to be
\begin{equation}
f_{p}^{\mathrm{IG}} \simeq 1.7 \times 10^3~\mathrm{Hz}\left(\frac{M_{\mathrm{in}}}{10^4 ~\mathrm{g}}\right)^{-5 / 6}\,\,.
\end{equation}
From Eq.~\eqref{omega4}, we can get the relation between the DM mass and the initial PBH mass,
\begin{equation}
    M_{\mathrm{in}} = 2.81\times 10^8~\mathrm{g} \left(\frac{M_\chi}{10^9~\mathrm{GeV}}\right)^{-2/5}\,\,,
\end{equation}
then we get,
\begin{equation}
    f_{{p}}^{\mathrm{IG}} \simeq 0.33~\mathrm{Hz} \left(\frac{M_\chi}{10^9~\mathrm{GeV}}\right)^{1/3}\,\,.
\end{equation}
Therefore, the peak frequency of induced GWs is directly linked to the DM mass, providing a novel approach for detecting DM. We show the results in Fig.~\ref{fig:ingw}. 
The left panel presents the spectrum of induced GWs. The right panel shows the contour plot for superheavy DM, along with the observational regions accessible to the GW experiments, like LISA, BBO, and ET.

\subsection{Gravitational waves from primordial black hole evaporation}

Even if PBHs never dominate the Universe, they can produce GWs through the direct evaporation of gravitons. The spectrum of these GWs can be expressed as follows:
\begin{equation}
    \frac{d \rho_{\mathrm{GW}}}{dt dp} = n_{\mathrm{PBH}}(t)p\frac{dN_{\mathrm{grav}}}{dtdp}\,\,.
\end{equation}
By using $n_{\mathrm{PBH}}(t) = n_{\mathrm{PBH}}(t_{\mathrm{in}})(a_{\mathrm{in}}/a)^3$, the GW spectra at $t_{\mathrm{evap}}$ can be written as
\begin{equation}
\frac{d \rho_{\mathrm{GW},\mathrm{evap}}}{d{\ln p_{\mathrm{evap}}}}=\frac{1}{ \pi^3} n_{\mathrm{PBH}}(t_{\mathrm{in}}) p_{\mathrm{evap}}^4 \int_{t_{\mathrm{in}}}^{t_{\mathrm{evap}}}dt \left(\frac{a_{\mathrm{in}}}{a}\right)^3\frac{\sigma_{s_i}^{l m}}{\exp \left[\left(p(t)-m \Omega\right) / T_{\mathrm{PBH}}\right]-1} \,\,.
\end{equation}
Here, we omit the summation of $l$ and $m$.
By using the relation
\begin{equation}
    p(t) =  p_{\mathrm{evap}}\frac{ a_{\mathrm{evap}}}{a} = p_0 \frac{a_0}{a} , \quad H(t) = H_{\mathrm{in}} \left(\frac{a_{\mathrm{in}}}{a}\right)^2 = \frac{\gamma M_{\mathrm{Pl}}^2}{2M_{\mathrm{in}}}\left(\frac{a_{\mathrm{in}}}{a}\right)^2\,\,,
\end{equation}
where $p_0$ is the momentum of graviton at present. Then we can get the GW energy spectrum at present,
\begin{equation}
\begin{aligned}
\frac{d \rho_{\mathrm{GW},0}}{d{\ln p_{0}}}&=\frac{1}{ \pi^3 H_{\mathrm{in}}} n_{\mathrm{PBH}}(t_{\mathrm{in}}) p_{0}^3 \frac{a_{\mathrm{in}}}{a_0}\int_{\frac{p_0 a_0}{a_{\mathrm{evap}}}}^{\frac{p_0 a_0}{a_{\mathrm{in}}}}dp \frac{\sigma_{s_i}^{l m}}{\exp \left[\left(p-m \Omega\right) / T_{\mathrm{PBH}}\right]-1} \\
& = \frac{1}{ \pi^3 H_{\mathrm{in}}} \frac{\beta \rho_{r}(T_{\mathrm{in}})}{M_{\mathrm{in}}} p_{0}^4 \frac{a_{\mathrm{in}}}{a_0}\int_{\frac{ a_0}{a_{\mathrm{evap}}}}^{\frac{ a_0}{a_{\mathrm{in}}}}dx \frac{\sigma_{s_i}^{l m}}{\exp \left[\left(xp_0-m \Omega\right) / T_{\mathrm{PBH}}\right]-1}\,\,,
\end{aligned}
\end{equation}
where in the second line we define $x = p/p_0$ and use Eq.~\eqref{beta}. Then we get
\begin{equation}
    \frac{d \rho_{\mathrm{GW},0}}{d{\ln p_{0}}}=2\times 10^{32} \beta M_{\mathrm{in}}^{-3/2} p_0^4 \int_{\frac{ a_0}{a_{\mathrm{evap}}}}^{\frac{ a_0}{a_{\mathrm{in}}}}dx \frac{\sigma_{s_i}^{l m}}{\exp \left[\left(xp_0-m \Omega\right) / T_{\mathrm{PBH}}\right]-1}\,\,,
\end{equation}
with
\begin{equation}
    \frac{ a_0}{a_{\mathrm{in}}} \simeq \left(\frac{g_{*s}(T_{\mathrm{in}})}{g_{*s}(T_{0})}\right)^{1/3}\frac{T_{\mathrm{in}}}{T_0}, \quad\frac{ a_0}{a_{\mathrm{evap}}} = \frac{ a_0}{a_{\mathrm{in}}} \frac{ a_\mathrm{in}}{a_{\mathrm{evap}}}\,\,.
\end{equation}
Determining the precise value of $a_{\mathrm{evap}}$ requires solving the Boltzmann equations of Kerr PBHs numerically, with or without superradiance. We use the public code BlackHawk~\cite{Arbey:2019mbc,Arbey:2021mbl} to get the greybody factors of Hawking radiation, which have the relation with the absorption cross section, $\Gamma(M_{\mathrm{PBH}},p) \equiv \sigma_{s_i}(M_{\mathrm{PBH}},p)p^2/\pi$. The relic abundance of GW from PBH evaporation is
\begin{equation}
    \Omega_{\mathrm{GW}}^{\mathrm{BH}}h^2 = \frac{h^2}{\rho_c}\frac{d \rho_{\mathrm{GW},0}}{d{\ln p_{0}}}\,\,.
\end{equation}

\begin{table}[t]
	\centering
	
	\setlength{\tabcolsep}{3mm}
	
	\begin{tabular}{c|c|c|c|c}
		\hline\hline
		& $M_\chi~\left(\mathrm{GeV}\right)$ & $M_{\mathrm{in}}~(\mathrm{g})$ &$a_\star^{\mathrm{in}}$ & $\beta$ \\
		\hline
        $BP_1$~(without SR)	 &$10^5$ & $10^8$ & 0.999 & $4.7\times 10^{-19}$  \\
            \hline
		$BP_2$~(with SR)	 &$10^5$ & $10^8$ & 0.999 & $1.3\times 10^{-20}$ \\
		\hline\hline
	\end{tabular}
	\caption{Two sets of parameters that satisfy the DM relic abundance with and without considering the superradiance.}\label{tab:ptable}
\end{table}

\begin{figure}[h]
	\centering
	\begin{minipage}{0.8\linewidth}
		\centering
		\includegraphics[width=0.9\linewidth]{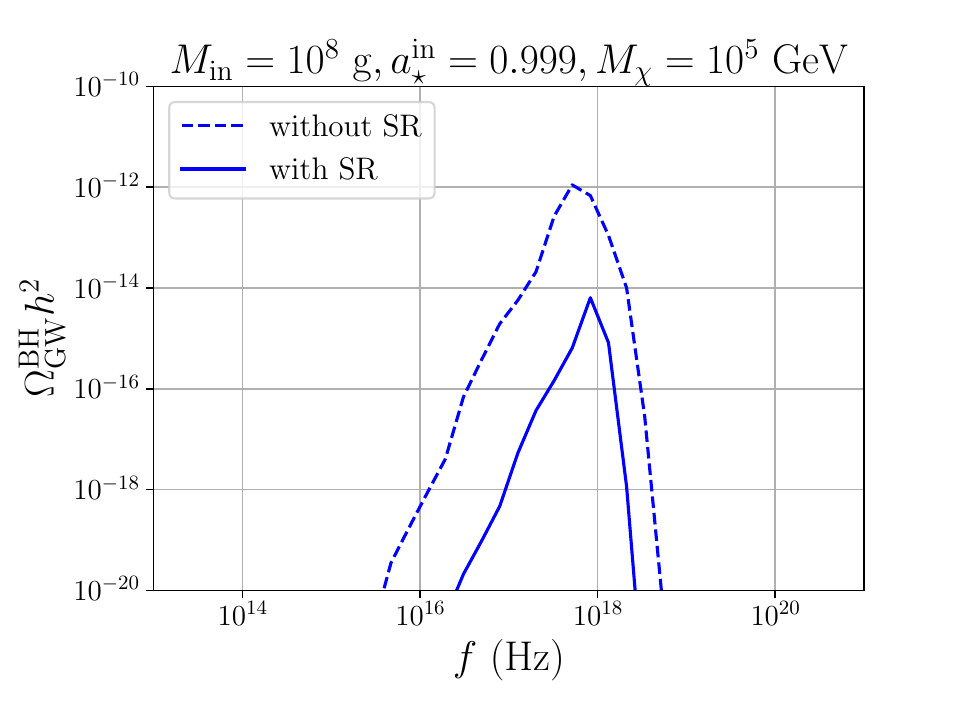}
	\end{minipage}
\caption{GW spectra generated by Hawking radiation. The solid and dashed lines correspond to the case with and without superradiance, respectively.}\label{fig:gwHW}
\end{figure}
The result of GW spectra are shown in Fig.~\ref{fig:gwHW}. The benchmark parameters are listed in Tab.~\ref{tab:ptable} where we set the same values of $M_\chi$, $M_{\mathrm{in}}$ and $a_\star^{\mathrm{in}}$. We can see that the effect of superradiance significantly alters the GW spectrum generated by Hawking radiation. This is because the superradiance process modifies the evolution of Kerr black holes in terms of their mass and angular momentum. Due to the appearance of superradiation, a large amount of energy and angular momentum is carried away by DM, so the strength of GW from Hawking radiation is significantly reduced. Although such extremely high-frequency GW spectra cannot yet be observed by current detectors, they may be detectable in the future, enabling us to effectively distinguish the contribution of superradiance.

\section{Possible interpretation of the KM3-230213A neutrino event from pNGB decay}\label{KM3}

Recently, the KM3NeT collaboration has reported the
detection of a high energy neutrino event, with an energy of $220_{-100}^{+570}$ PeV~\cite{KM3NeT:2025npi}. This ultra-high energetic neutrino signals may point towards some  astrophysical interpretations~\cite{KM3NeT:2025aps} or cosmic origin~\cite{KM3NeT:2025vut,Boccia:2025hpm,Borah:2025igh,Kohri:2025bsn,Narita:2025udw,Barman:2025bir,Brdar:2025azm,Wu:2023twu}. In this work, we consider the origin of this neutrino event from pNGB decay.

We use the models in Refs.~\cite{Abe:2020iph,Okada:2020zxo}.
which introduce the gauged $U(1)_{B-L}$ symmetry. The $S$ carries a charge of $Q_{B-L} = +1$. Additionally, we introduce a complex scalar field $\Phi = (v_\phi+\phi)e^{i\eta}/\sqrt{2}$ with $Q_{B-L} = +2$.
The potential of the UV model enlarges Eq.~\eqref{potential} as,
\begin{equation}
    \begin{aligned}
V(H, S, \Phi)=&-\frac{\mu_H^2}{2}|H|^2-\frac{\mu_S^2}{2}|S|^2-\frac{\mu_{\Phi}^2}{2}|\Phi|^2+\frac{\lambda_H}{2}|H|^4+\frac{\lambda_S}{2}|S|^4+\frac{\lambda_{\Phi}}{2}|\Phi|^4\,\,,\\
&+\lambda_{H S}|H|^2|S|^2+\lambda_{H \Phi}|H|^2|\Phi|^2+\lambda_{S \Phi}|S|^2|\Phi|^2-\left(\frac{\mu_{\mathrm{c}}}{\sqrt{2}} \Phi^* S^2+\text{h.c.}\right) \,\,.
\end{aligned}
\end{equation}
We set $\lambda_{H\Phi} = \lambda_{S\Phi} = 0$ for simplicity. Due to the last term, after the symmetry breaking of $\Phi$ we can get the explicit breaking term in Eq.~\eqref{potential}. The $\Phi$ can couple to the right-handed neutrinos through the term $f \Phi \overline{\nu_{R}^c} \nu_{R}$.
The symmetry breaking of $\Phi$ gives the right-handed neutrino a mass of $M_N = f v_\phi / \sqrt{2}$. Additionally, the electroweak symmetry breaking of the SM Higgs field $H$ provides a Dirac mass for neutrinos, $m_D = y^\nu v / \sqrt{2}$. Thus, based on the standard type-I seesaw mechanism, the neutrino mass is given by $\left(m_\nu\right)_{\alpha \beta} \approx \left(m_D M_N^{-1} m_D^{\mathrm{T}}\right)_{\alpha \beta}$.
Due to the final mixing term, $\chi$ and $\eta$ can mix with each other, with the mixing angle being proportional to $v_s/v_\phi$. One of the NGBs is absorbed by the dark gauge boson $Z'$, while the other remains as a physical pNGB. The pNGB can decay into neutrinos through NGB mixing and active-sterile neutrino mixing. When $v_s/v_\phi \ll 1$, this decay width is much smaller than that of the Majoron, which is given by Eq.~\eqref{decaywidth}. Additionally, because we only consider the case $M_\chi < M_s$, the pNGB can decay into SM fermions through the off-shell $s$ and $Z'$, $\chi \rightarrow s Z' \rightarrow \overline{f}_{\mathrm{SM}}f_{\mathrm{SM}}\overline{f}_{\mathrm{SM}}f_{\mathrm{SM}}$~\cite{Okada:2020zxo}. The decay width is proportional to $\frac{M_\chi^9}{v_\phi^4 M_s^4}$ and $\sin \theta_{sh}^2$ where $\sin \theta_{sh}$ is the mixing between the $s$ and the SM Higgs. Therefore this decay channel may dominate when $M_\chi$ is large and the mixing angle is non-zero. Note that the mixing angle also serve as the origin of freeze-in production of pNGB DM. We neglect the kinetic mixing of the gauge bosons in this analysis.

In order to explain the high-energy neutrino event, we set $M_\chi = 440~\mathrm{PeV}$. For such superheavy DM, traditional production mechanism like freeze-out is hard to work. However, in this work we have shown that the DM could be directly produced by black holes. We consider the decay channels of pNGB  primarily into neutrinos and top quarks, and their secondaries such as gamma-rays and high energy neutrinos which could explain the event of KM3NeT. The neutrino flux can be estimated as
\begin{equation}
    \frac{d^2\Phi_\nu^G}{d E_\nu d\Omega} = \frac{1}{\tau_\chi} \frac{\mathcal{D}}{4\pi M_\chi}\frac{dN_\nu}{d E_\nu }\,\,,
\end{equation}
where $dN_\nu/dE_\nu$ is the neutrino energy spectrum and the $\mathcal{D}$-factor
is defined as
\begin{equation}
\mathcal{D}=\frac{1}{\Delta \Omega} \int_{\Delta \Omega} d \Omega \int_0^{s_{\max }} d s \rho\left(r(s,\psi)\right)\,\,,
\end{equation}
where $r(s,\psi) = \sqrt{s^2 + r_\odot^2 - 2sr_\odot \cos\psi}$ is the radial coordinate, $\cos \psi = \cos b\cos l $ with $(b, l)$ being the galactic coordinates, $d \Omega=d \sin b~d l$ and $\Delta \Omega= \int_l \int_b d \Omega$, $r_\odot \simeq 8.5~\mathrm{kpc}$ is  the
 distance from the center of the Milky Way to the solar system. And $s_{\max }=\sqrt{d_{\mathrm{MW}}^2-r_\odot^2\sin ^2 \psi }+r_\odot \cos \psi$, 
 where $d_{\mathrm{MW}}=40~\mathrm{kpc}$ is the diameter of the Milky Way. The DM density profile $\rho(r)$ in the Milky Way can be given as
\begin{equation}
\rho(r)=\frac{\rho_c}{\left(\frac{r}{r_s}\right)^\gamma\left[1+\left(\frac{r}{r_s}\right)^\alpha\right]^{(\beta-\gamma) / \alpha}}\,\,,
\end{equation}
where $r_s$ is the scale radius and $\alpha, \beta$, and $\gamma$ are slope parameters. For the Navarro-Frenk-White (NFW) profile~\cite{Navarro:1996gj,Akita:2022lit} $\alpha=1$, $\beta=3$, $ \gamma=1$, $r_s=20~\mathrm{kpc}$, and we set $\rho_c$ such that $\rho\left(r_\odot\right)=0.4 ~\mathrm{GeV~cm}^{-3}$.

\begin{figure}[h]
	\centering
	\begin{minipage}{0.8\linewidth}
		\centering
		\includegraphics[width=0.9\linewidth]{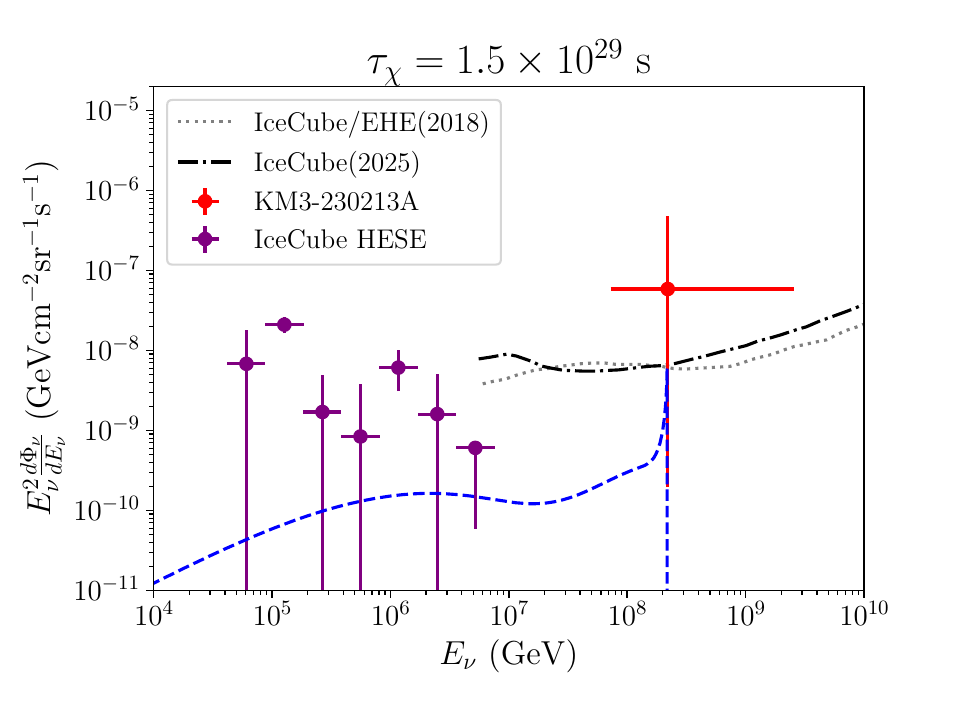}
	\end{minipage}
\caption{Neutrino flux (the blue dashed line) from pNGB DM decay. The red data point shows the high-energy neutrino event. The purple data points show the IceCube’s HESE events. The gray dotted line and the black dashed dotted line correspond to  IceCube-EHE data and 12.6 years of IceCube data, respectively.}\label{fig:nflux}
\end{figure}

The extragalactic neutrino flux resulting from the decay of pNGB DM is~\cite{Blanco:2018esa}
\begin{equation}
\begin{aligned}
\frac{d \Phi_{\nu}^{\mathrm{EG}}}{d E_\nu}= \frac{1}{\tau_\chi} \frac{1}{4\pi M_\chi} \int_0^{\infty} dz \frac{\rho_0 c / H_0}{\sqrt{\Omega_m(1+z)^3+\Omega_{\Lambda}}} \frac{1}{1+z}  \left(\frac{d N_{\nu}}{d E'_{\nu}}\right)_{E'_{\nu} = (1+z)E_\nu}\,\,,
\end{aligned}
\end{equation}
where $z$ is the redshift and $\rho_0=1.15 \times 10^{-6}~\mathrm{GeV~cm}^{-3}$ is the present cosmological average DM density. $\Omega_m=0.315$ and  $\Omega_{\Lambda}=0.685$~\cite{Planck:2018vyg} are the energy fraction of matter and vacuum energy, respectively. $c / H_0=$ $1.37 \times 10^{28} \mathrm{~cm}$ is the proper Hubble radius. Then the total flux per unit solid angle is
\begin{equation}
\frac{d\Phi_{\nu}}{dE_\nu} \equiv \frac{d^2 \Phi_{\nu}^{\mathrm{G}}}{d E_{\nu} d \Omega}+\frac{1}{4 \pi} \frac{d \Phi_{\nu}^{\mathrm{EG}}}{d E_{\nu}}
\end{equation}
We obtain the neutrino spectrum $dN_\nu/dE_\nu$ by using the  publicly available code
HDMSpectra~\cite{Bauer:2020jay}. We take the life time of DM $\tau_\chi$ as a free parameter.
Also it is assumed that due to the neutrino oscillations, the neutrinos have the flavor ratio 1:1:1 when reaching the Earth.

The non-observation of the high-energy neutrino event by experiments like IceCube sets the upper limit on the high-energy neutrino flux. The predicted neutrino flux is shown in terms of the blue dashed line in Fig.~\ref{fig:nflux}. It can be seen that the flux fits the observation of KM3NeT well.
The purple data points show the IceCube’s HESE events~\cite{IceCube:2020wum}. The gray dotted line and the black dashed dotted line correspond to  IceCube-EHE data~\cite{IceCube:2018fhm} and 12.6 years of IceCube data~\cite{IceCube:2025ezc}, respectively. 
The lifetime of DM must exceed \( 1.5 \times 10^{29} \) s to remain consistent with IceCube constraints.
This also tells us why the Majoron cannot account for this high-energy signal. The Majoron’s lifetime is not sufficiently long for $v_s \lesssim M_{\mathrm{Pl}}$.

\section{Conclusions and discussions}\label{conclusions}

Due to the nature of NGB, the pNGB DM as a compelling DM candidate, is unlikely to be produced via the traditional thermal freeze-out mechanism and is instead more plausibly generated through alternative non-thermal processes. Thus, in this work, we have explored novel production mechanisms for pNGB DM via Hawking radiation and superradiance from PBHs，as well as the associated GW signatures.

Our study encompasses the production of both light and superheavy pNGB DM under the influence of Schwarzschild and Kerr black hole scenarios.
For Schwarzschild black holes, we derived analytical approximate solutions for the DM production. We investigate the effects of Hawking radiation and UV freeze-in, respectively. In the case of Kerr black holes, we numerically solved the evolution differential equations to determine the DM relic abundance generated by Hawking radiation and superradiance. 

Additionally, we explored the GW signals linked to these processes, identifying detectable signatures from DW annihilation and the induced GWs, which are within the sensitivities of experiments such as LISA, TianQin, Taiji, BBO, ET, CE, and other proposals. Furthermore, high-frequency GWs resulting from PBH evaporation may be accessible to future high-frequency GW detection experiments.

Finally, we proposed that in a viable UV completion model, the pNGB decay could account for the recently reported $\mathcal{O}$(100) PeV neutrino event KM3-230213A at KM3NeT. The pNGB decays through the scalar mixing and is long-lived enough to avoid the constraint from the experiments like IceCube.

These findings highlight the interplay between black hole physics, DM production, and multi-messenger signals, emphasizing the relevance of forthcoming experimental advancements in probing the fundamental nature of DM.

\tc{We have assumed a monochromatic distribution for the mass and spin of PBHs in this work. However, more realistic mass and spin distributions may have significant implications for cosmology. For instance, a log-normal mass distribution—commonly predicted in slow-roll inflation scenarios~\cite{Dolgov:1992pu,Carr:2016drx,Green:2016xgy} can affect PBH evolution through its width~\cite{Cheek:2022mmy}. A broader distribution leads to a more prolonged evaporation process, thereby modifying the resulting DM relic abundance and the GW spectrum associated with PBH evaporation. Extending our analysis to include PBH superradiance with general mass and spin distributions is a worthwhile direction, which we leave for future work. Moreover, the cosmological origin of Kerr PBHs with near-extremal spins remains uncertain and warrants further investigation. Another intriguing avenue involves exploring the memory burden effects of PBHs, which influence their evolution and may have important consequences for DM production~\cite{Haque:2024eyh,Barman:2024iht,Borah:2024bcr,Basumatary:2024uwo,Zantedeschi:2024ram}.}

\acknowledgments
We thank Yuber F. Perez-Gonzalez and Andrew Cheek for the helpful guidance about their code ULYSSES. We thank Yong Xu for bringing our attention to the necessity of considering gravitational production and Jongkuk Kim for reminding us of the direction of the neutrino flux. This work is supported by the National Natural Science Foundation of China (NNSFC) under Grant No.12205387 and No.12475111.
\end{CJK}

\bibliographystyle{JHEP}
\bibliography{majorogenesis}


\end{document}